\documentclass[apj]{emulateapj}
\usepackage{natbib}
\usepackage{graphicx}
\usepackage[toc,page]{appendix}
\usepackage{natbib}
\usepackage{color}
\usepackage{threeparttable}
\usepackage{hyperref}
\usepackage{breakurl}
\makeatletter

\newcommand{\Rmnum}[1]{\expandafter\@slowromancap\romannumeral #1@}
\makeatother

\begin{document}

\slugcomment{Accepted for publication in the Astrophysical Journal}
\title{Resolved Star Formation on Sub-galactic Scales in a Merger at $z=1.7$}
\email{kate.whitaker@nasa.gov}
\author{Katherine E. Whitaker\altaffilmark{1,6}, Jane R. Rigby\altaffilmark{1}, Gabriel B. Brammer\altaffilmark{2},
Michael D. Gladders\altaffilmark{3}, \\
Keren Sharon\altaffilmark{4}, Stacy H. Teng\altaffilmark{1,6}, Eva Wuyts\altaffilmark{5}}
\altaffiltext{1}{Astrophysics Science Division, Goddard Space Flight Center, Code 665, Greenbelt, MD 20771, USA}
\altaffiltext{2}{Space Telescope Science Institute, 3700 San Martin Drive, Baltimore, MD, USA}
\altaffiltext{3}{The Department of Astronomy and Astrophysics, and the Kavli Institute for Cosmological Physics, 
  The University of Chicago, 5640 South Ellis Avenue, Chicago, IL 60637, USA}
\altaffiltext{4}{Department of Astronomy and Astrophysics, University of Michigan, 500 Church Street, 
  Ann Arbor, MI 48109, USA}
\altaffiltext{5}{Max-Planck-Institut f\"{u}r extraterrestrische Physik (MPE), Giessenbachstr., D-85748 Garching, Germany}
\altaffiltext{6}{NASA Postdoctoral Program Fellow}

\shortauthors{Whitaker et al.}
\shorttitle{Resolved Star Formation on Sub-galactic Scales}

\begin{abstract}
We present a detailed analysis of Hubble Space Telescope (HST), Wide Field Camera 3 (WFC3) G141 
grism spectroscopy for seven star-forming regions 
of the
highly magnified lensed starburst galaxy RCSGA 032727-132609 at $z=1.704$.  We measure the spatial variations of the extinction in 
RCS0327 through the observed H$\gamma$/H$\beta$ emission 
line ratios, finding a constant average extinction of $\mathrm{E(B-V)_{gas}}=0.40\pm0.07$.  
We infer that the star formation is enhanced as a result of an ongoing interaction, with measured star formation
rates derived from demagnified, extinction-corrected H$\beta$ line fluxes for the individual star-forming clumps falling 
$>$1--2 dex above the star formation sequence.
When combining the HST/WFC3 $\mathrm{[OIII]}\lambda5007$/H$\beta$ emission line ratio measurements 
with [NII]/H$\alpha$ line ratios from \citet{EWuyts14}, we find that the majority of 
the individual star-forming regions fall along the local ``normal'' abundance sequence.  
With the first detections of the He~I $\lambda$5876 $\mathrm{\AA}$ and He~II $\lambda$4686 $\mathrm{\AA}$ 
recombination lines in a distant galaxy, we
probe the massive-star content of the star-forming regions in RCS0327.  The majority of the star-forming 
regions have a He~I $\lambda$5876 to H$\beta$ ratio consistent with the saturated maximum value, 
which is only possible if they still contain hot O-stars.  Two
regions have lower ratios, implying that their last burst of new star formation ended $\sim5$ Myr ago.  
Together, the He~I $\lambda$5876 $\mathrm{\AA}$ and He~II $\lambda$4686 $\mathrm{\AA}$ to H$\beta$ line ratios
provide indirect evidence for the order in which star formation is stopping in individual star-forming knots of 
this high redshift merger.
We place
the spatial variations of the extinction, star formation rate and ionization conditions in the context of 
the star formation history of RCS0327. 
\end{abstract}


\section{Introduction}
\label{sec:intro}

Detailed studies of galaxies only a few billion years after the Big Bang are critical for our 
understanding of galaxy formation.  During this epoch, the majority of stars in the Universe 
were formed \citep{Hopkins06} and central supermassive black holes were most active \citep{PHopkins08}.
Our knowledge of star formation in distant galaxies at `cosmic noon', or $z$$\sim$1--2, is largely drawn from global
averages over marginally resolved galaxies.  Despite the inherent challenges, 
these studies have enabled us to trace galaxy correlations
across billions of years to reconstruct their formation histories \citep[e.g.,][and many more]{Marchesini09,Whitaker12b,
Bell12,Cassata13,Tomczak13}.  

However, the internal dynamics and kinematics of galaxies are 
known to be more complicated.  Galaxies exhibit gradients in their stellar populations, such as age, color and metallicity
\citep[][and numerous others]{Bell00,MacArthur04,Ferreras05,Szomoru13,SanchezBlazquez13}. Furthermore,
measurements of global extinctions vary widely amongst different star-forming 
populations \citep[e.g.,][]{Erb06,Takata06,Garn10,Dominguez13,Zahid13,Kriek13}. 
Although we would ideally like to resolve the spatial variations of stellar populations on sub-galactic
scales for distant galaxies, we often rely upon integrated colors and spectroscopy.
This issue of spatial resolution only becomes more
challenging as we push studies to earlier epochs.  

\begin{figure*}[t]
\leavevmode
\centering
\includegraphics[width=0.9\linewidth]{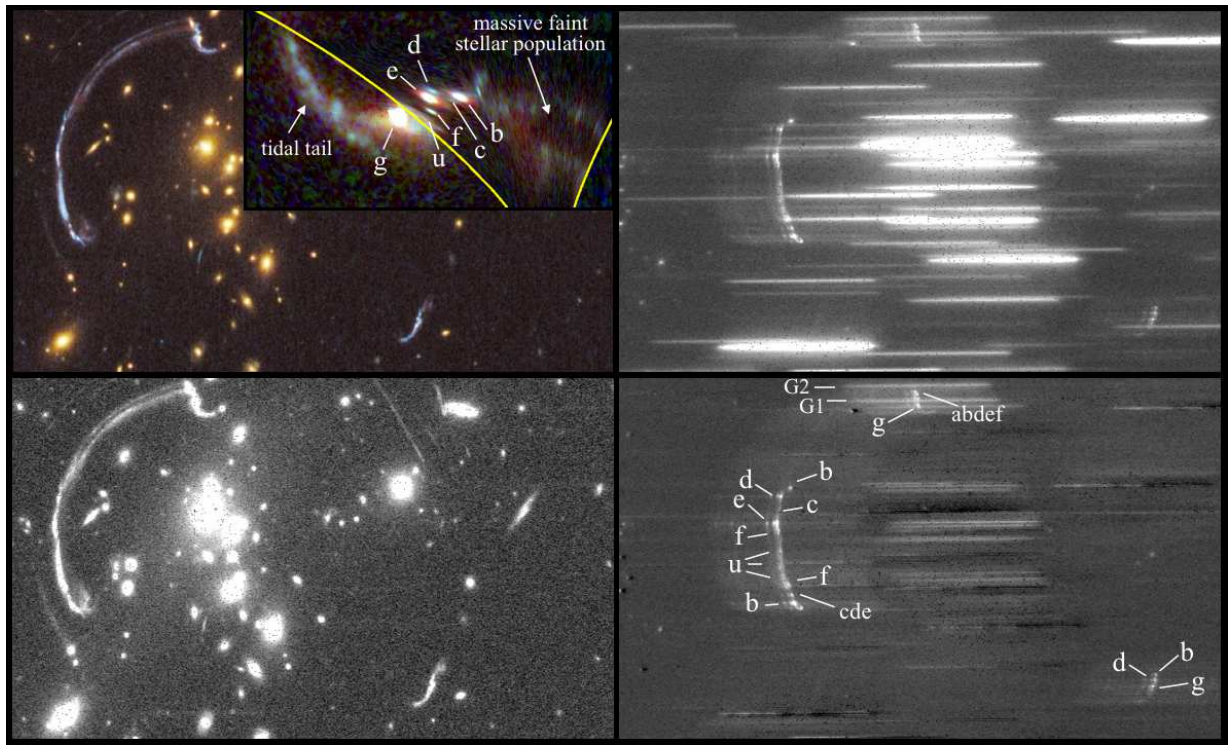}
\caption{HST/WFC3 grism spectroscopy of the lensed star-forming galaxy RCS0327 at $z=1.704$ reveals the emission-line
properties of multiple star-forming regions.
The top left panel and inset show the color rendition of the giant arc
on-sky and the source reconstruction for RCS0327 from \citet{Sharon12},
composed of F390W (blue), F606W+F814W (green), and F098M+F125W+F160W (red).
In this paper, we introduce the direct HST/WFC3 F140W image (bottom left), raw G141 grism spectra (top right), and
the grism spectra after subtracting the foreground galaxy cluster contamination (bottom right).  
All fluxes shown here are not demagnified.}
\label{fig:image}
\end{figure*}

Although fundamental for probing global galaxy correlations across cosmic time, 
it is not clear if the integrated stellar populations necessarily encapsulate variations of physical properties on 
the smaller scales of star-forming regions.  
Gravitational lensing enables astronomers to probe new parameter space that would not otherwise
be accessible with the current generation of telescopes.
High redshift galaxies are magnified such that we
can study distant samples to lower luminosities, and, for the brightest lensed sources, we can even resolve 
individual star forming clumps at high spatial resolution.
Lensed galaxies are the ideal laboratories
for studying the (spatially-resolved) properties of distant galaxies \citep[e.g.,][]{Jones10,EWuyts10,Richard11,Stark13}.  
For example, \citet{Yuan11} measured the first
metallicity gradient for a lensed grand-design face-on spiral galaxy at $z\sim1.5$, finding well-developed spiral arms
and nebular emission line dynamics indicative of a rotationally supported disk that has a significantly steeper
metallicity gradient than local galaxies.

\begin{figure*}[t]
\leavevmode
\centering
\includegraphics[width=0.8\linewidth]{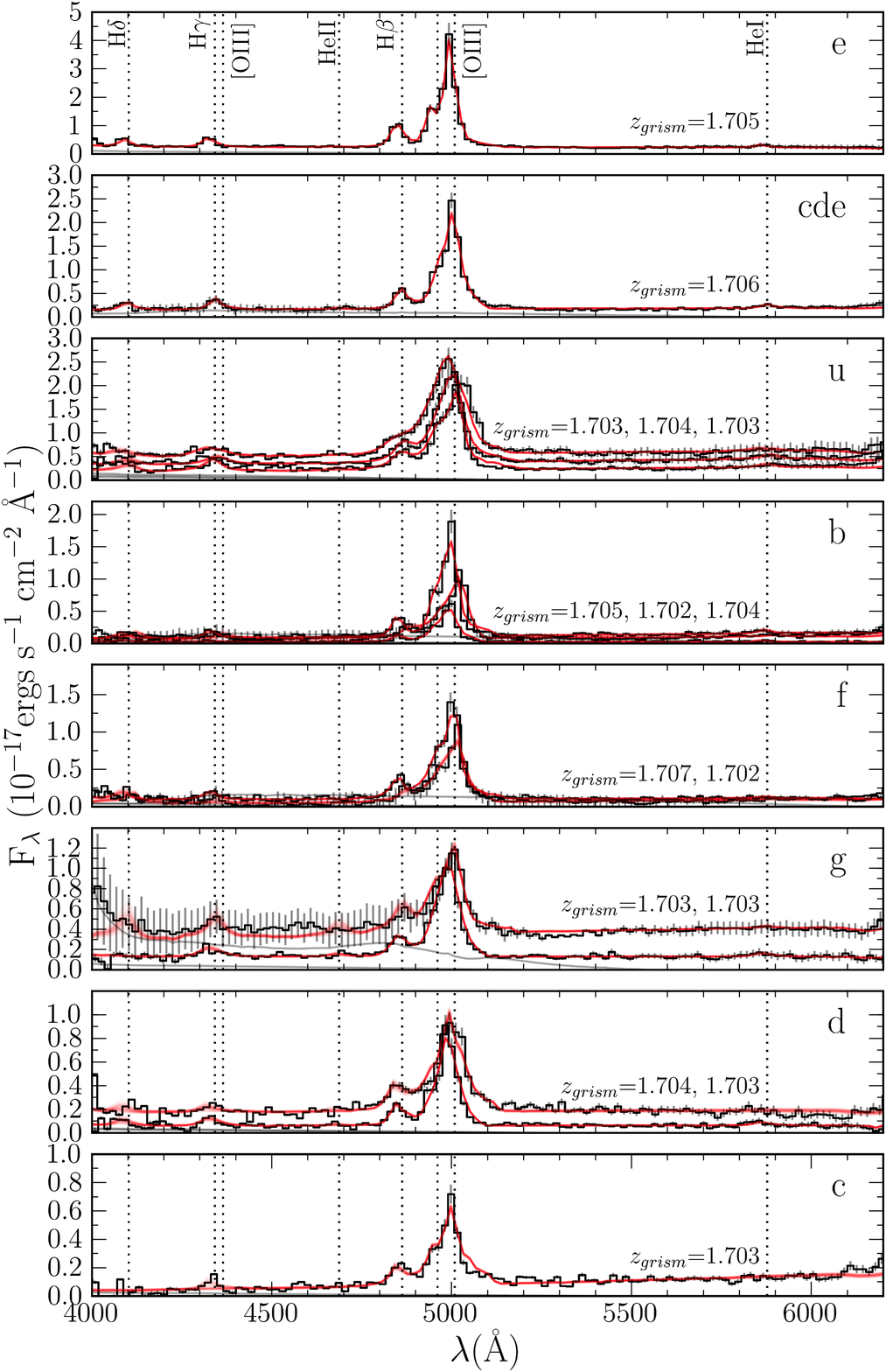}
\caption{One dimensional grism spectra for individual star-forming regions, as labeled in Figure~\ref{fig:image}.
The grism redshifts agree well with the spectroscopic redshift of $z_{spec}=1.704$.
Several clumps have multiple images, each showing different morphological broadening.  The
red lines indicate the best-fit convolved model spectra from a Markov chain Monte Carlo analysis,
and the contamination is shown in grey.}
\label{fig:spectra}
\end{figure*}

Among the brightest known distant lensed galaxies is RCSGA 032727-132609 \citep[][hereafter RCS0327]{EWuyts10}. With a redshift of $z=1.704$, 
it is one of the best-studied star-forming galaxies at the peak of cosmic star formation.  Due to both the strong
gravitational lensing magnification and enhanced star formation from an ongoing interaction, RCS0327 provides a
unique opportunity to study a distant galaxy in detail on the $\sim100$ pc spatial scales characteristic of star-forming regions.
Through a kinematic analysis of adaptive optics (AO) assisted integral field spectroscopic data using the
OH-Suppressing Infra-Red Imaging Spectrograph (OSIRIS), \citet{EWuyts14} find 
that RCS0327 is an ongoing interaction with a large tidal tail.  Four of the seven identified individual 
star-forming regions from Hubble Space Telescope (HST) Wide Field Camera 3 (WFC3) 
imaging (diameters ranging from 300 to 600 pc) were observed with high enough
signal-to-noise (S/N) to show further evidence 
for outflows from the H$\alpha$ emission line profiles.  

The ongoing interaction within RCS0327 boosts the global specific star formation rate to a factor of three 
above the main sequence of star-formation at $z\sim2$ \citep[e.g.,][]{Whitaker12b}.  The nature of such outliers on the 
star formation sequence are not well understood \citep[e.g.,][]{Dave12}.  Two possible modes of star formation 
that control the growth of galaxies include a relatively steady mode for disk-like galaxies \citep[e.g.,][]{Peng10} and a starburst mode, 
generally attributed to merger events \citep[e.g.,][]{Mihos94,Bournaud11}.  
\citet{Rodighiero11} find that starburst galaxies represent only 2\% of mass-selected star-forming galaxies at the peak
epoch of star formation activity, and thus merger-driven starbursts play a relatively minor role in the formation of
stars in galaxies.
Here, we can compare the physical properties of the ongoing merger RCS0327 to understand the spatial variations in the 
star formation activity enhancements.  In the future, larger samples of distant lensed galaxies can shed light on
the primary mechanism governing the build-up of stars within these galaxies.

In this paper, we present spatially-resolved H${\gamma}$/H${\beta}$, $\mathrm{[O\Rmnum{3}]}\lambda5007$/H${\beta}$ and 
$\mathrm{He\Rmnum{1}}\lambda5876$/H${\beta}$ line ratios from HST/WFC3 grism observations of RCS0327.  We measure the average
absolute extinction and map the variation in extinction across the giant arc in Section~\ref{sec:dust}.  We further
map the star formation rate and ionization conditions within the galaxy in Sections~\ref{sec:sfr} and \ref{sec:bpt}, respectively.  
In particular, we present the first spatially-resolved
measurements of the Helium to Hydrogen recombination line ratios for a star-forming galaxy at $z>1$ in Section~\ref{sec:HeI}.  

We assume a $\Lambda$CDM cosmology with $\Omega_{M}$=0.3, $\Omega_{\Lambda}$=0.7, $H_{0}$=70 km s$^{-1}$ Mpc$^{-1}$ 
and a \citet{Kroupa01} initial mass function (IMF) throughout the paper.  All magnitudes are 
given in the AB system.

\section{Observations}
\label{sec:observations}

Observations with the HST/WFC3 G141 grism were taken over two epochs
on February 13, 2012 and February 21, 2013 under GO Program 12726 (PI: Rigby).  During each visit, the direct image of RCS0327 
was observed with the F140W filter (228 s), along with the dispersed light from the G141 grism (2603 s) for two dither positions.
To minimize crowding from cluster galaxies and enable cross-checks on the measurements, the field was observed
at a different roll angle for each visit separated by 5.5$^{\circ}$.  We find consistent results between
both epochs, and co-add the data for the analyses presented herein.

Figure~\ref{fig:image} shows the color image and spectrum for the lensed galaxy in RCS0327. 
The HST/WFC3 F140W image is shown in the bottom left panel with the raw G141 grism observations in the top right panel.
After subtracting the contaminating light from the foreground cluster galaxies as described in 
Section~\ref{sec:analysis}, we label the star-forming clumps on the ``clean'' G141 grism image presented herein (bottom right panel).  
The spectra were reduced as described in detail by \citet{Brammer12}.  

\begin{table*}[ht]
\centering
\begin{threeparttable}
    \caption{Measured line fluxes from HST/WFC3 G141 grism spectroscopy}\label{tab:data}
    \begin{tabular}{lccccc}
      \hline \hline
      \noalign{\smallskip}
      Clump & H$\gamma$+$[\mathrm{O{\Rmnum{3}}}]\lambda4363$ & $\mathrm{He\Rmnum{2}}\lambda4686$ & H$\beta$ & $[\mathrm{O{\Rmnum{3}}}]\lambda4959+5007$ & $\mathrm{He\Rmnum{1}}\lambda5876$  \\
      \noalign{\smallskip}
      \hline
      \noalign{\smallskip}
      \emph{b}  & $6.1\pm3.5$ & -- & $9.0\pm1.8$ & $90.5\pm1.8$ & $1.8\pm1.2$ \\
                & $5.6\pm1.3$ & -- & $15.5\pm0.8$ & $114.2\pm1.0$ & $3.1\pm1.0$ \\
                & -- & -- & $5.5\pm0.6$ & $39.4\pm0.7$ & $0.9\pm0.4$ \\
      \emph{c}  & -- & -- & $9.4\pm1.1$ & $50.3\pm1.4$ & -- \\
      \emph{d}  & $3.8\pm2.2$ & -- & $18.3\pm2.1$ & $79.8\pm2.5$ & -- \\
                & $4.3\pm0.8$ & -- & $10.9\pm0.5$ & $61.0\pm0.6$ & $1.9\pm0.5$ \\
      \emph{e}  & $16.7\pm1.3$ & $1.0\pm0.7$ & $41.7\pm1.1$ & $227.5\pm1.2$ & $4.3\pm1.3$ \\
      \emph{g}  & -- & -- & -- & $84.6\pm2.7$ & -- \\
                & $6.5\pm1.1$ & $1.6\pm0.9$ & $14.1\pm0.7$ & $80.0\pm0.7$ & $2.5\pm0.7$ \\
      \emph{u}  & $19.5\pm2.5$ & -- & $39.3\pm1.6$ & $246.3\pm1.9$ & $5.6\pm3.6$ \\
                & $14.7\pm2.4$ & -- & $29.6\pm1.4$ & $191.5\pm1.7$ & $8.0\pm3.1$ \\
                & -- & -- & $28.2\pm1.3$ & $163.6\pm1.5$ & -- \\
      \emph{cde}& $11.3\pm1.8$ & -- & $22.1\pm1.3$ & $147.1\pm1.5$ & $4.2\pm0.9$ \\
      \noalign{\smallskip}
      \hline
      \noalign{\smallskip}
    \end{tabular}
    \begin{tablenotes}
      \small
    \item \emph{Notes.} Fluxes are not demagnified and are in units of 10$^{-17}$ erg s$^{-1}$ cm$^{-2}$.
    \end{tablenotes}
  \end{threeparttable}
\end{table*}

The top left panel of Figure~\ref{fig:image} shows the on-sky color rendition of the giant arc from \citet{Sharon12}, including 6
HST filters: F390W (blue), F606W+F814W (green), and F098M+F125W+F160W (red).
The inset panel is the reconstruction of the source from
the lens model described in \citet{Sharon12}.  Stellar mass surface density maps suggest that the stellar populations
in clump \emph{g} were established earlier than the other star-forming clumps, and clump \emph{g} is merging with an old, 
red (faint) stellar population at the right side of the reconstructed
image \citep{EWuyts14}.  The bright star-forming knots in between are likely a result of the interaction, as demonstrated herein.

\section{Data Analysis}
\label{sec:analysis}

Figure~\ref{fig:spectra} shows the extracted one-dimensional (1D) spectra of seven individual star-forming knots within
the bright lensed galaxy RCS0327.  The individual star-forming regions (clumps \emph{b, c, d, e, f, g}, and 
\emph{u}) are labeled following the definitions of \citet{EWuyts10}.
Strong emission lines of $[\mathrm{O{\Rmnum{3}}}]\lambda4959+5007$
and H${\beta}$, as well as the fainter H${\gamma}$ (blended with $[\mathrm{O{\Rmnum{3}}}]\lambda4363$), H${\delta}$,
$\mathrm{He\Rmnum{2}}\lambda4686$ and $\mathrm{He\Rmnum{1}}\lambda5876$ emission lines are detected.
Table~\ref{tab:data} lists the measured emission line fluxes for the unique star-forming clumps.
Due to the slitless configuration of the HST/WFC3 grism, the lines are broadened following the two-dimensional (2D)
morphology of the source.
Where multiple images are available, the
effects of morphological broadening and magnification become evident from the varying line profiles and              
line flux strengths of the same physical
regions in Figure~\ref{fig:spectra}.  In particular, clumps \emph{u}, \emph{b} and \emph{f}
show cases where the dispersion axis is perpendicular to the arc (resulting in symmetric line profiles) and
at a slight angle with respect to the dispersion axis (asymmetric line profiles).

We include only grism spectra where we are able to
unambiguously isolate individual physical regions, with
the exception of the blended spectrum for clumps \emph{c}, \emph{d} and \emph{e} due to the high S/N detection
of $\mathrm{He\Rmnum{1}}\lambda5876$.  Clump \emph{e} only has one unique spectrum, as all other images of this physical region are blended
with other physical regions.  The image overlapping with ``pointing 3'' of \citet{EWuyts14} contains two bright cluster galaxies,
labeled \emph{G1} and \emph{G2} in Figure~\ref{fig:image}.  These cluster galaxies highly contaminate the spectrum of clump \emph{g} in 
this region, as seen through the large error bars and contamination (grey line) in Figure~\ref{fig:spectra},
rendering these data unusable.  Although we include all extracted spectra that uniquely correspond to an individual
star-forming region in Figure~\ref{fig:spectra}, we disregard this single observation of clump {g}, as well as one of
\emph{u} and clump \emph{b}
for the remainder of the analyses due to large uncertainties from morphological broadening and contamination.

In order to properly extract line fluxes from the grism spectrum, we generate and fit a model to
the 2D spectrum following \citet{Brammer12b}.  Next, we briefly summarize the main steps.
The contamination from cluster galaxies is determined empirically from the observed F140W image.
To generate the contamination model spectra to be subtracted, we take the observed F140W flux and full
spatial profiles within each given object's segmentation map and
assume a flat spectrum in units of f$_{\lambda}$ and allow for a tilt correction.
We define extraction regions following the physical regions identified in \citet{Sharon12}.
We convolve the observed (contamination-subtracted) F140W arc morphology in each 
extraction region with a 1D spectrum that consists of a flat continuum plus individual emission lines to determine
the redshift.  Further details about the redshift fitting method can be found in Section 4.2 of \citet{Brammer12}.  

Once the redshift is established, the individual emission lines are fit.
The model is fit to the observed 2D spectrum with parameter optimization using the Markov chain Monte Carlo (MCMC)
sampler \citep{ForemanMackey13}.  The free parameters include the normalization and slope of a flat continuum to improve
the background subtraction and the individual
strengths of the emission lines $\mathrm{He\Rmnum{1}}\lambda5876$, $\mathrm{[O\Rmnum{3}]}\lambda4959+5007$, H${\beta}$, 
$\mathrm{He\Rmnum{2}}\lambda4686$, H${\gamma}+\mathrm{[O\Rmnum{3}]}\lambda4363$ and H${\delta}$. 
While the spectral resolution of the G141 grism ($\mathrm{R}\sim130$) is insufficient to resolve
the H${\gamma}$ and $\mathrm{[O\Rmnum{3}]}\lambda4363$ lines, the sum of these lines is well
constrained by the G141 spectrum \citep[see also][]{Brammer12b}.  When fitting the sum of these lines, we assume that the 
$\mathrm{[O\Rmnum{3}]}\lambda4363$ line is 10\% the flux of H${\gamma}$, as derived from H$\gamma$ and the upper limit of 
$\mathrm{[O\Rmnum{3}]}\lambda4363$ measured with higher resolution 
Near Infrared Spectrometer (NIRSPEC) data of clumps \emph{e} and \emph{u} \citep{Rigby11}.  The red lines in Figure~\ref{fig:spectra} show 100 samples from the 
MCMC chain.  In the following sections, we interpret the measured line fluxes within the individual star-forming
regions to measure the variation in the physical conditions within RCS0327.

\section{Mapping Extinction Variations}
\label{sec:dust}

The nature of dust in distant galaxies is not well understood: when 
modeling the spectral energy distributions of galaxies, one
assumes a universal dust extinction law to describe how dust absorbs and 
re-emits the stellar light of galaxies that exhibit a vast range in their star
formation histories and environments.  Although still poorly understood, 
it is clear that the dust attenuation law is not universal \citep[e.g.,][]{Stecher65,
Calzetti00, Weingartner01} and is sensitive to both 
the dust geometry and viewing angle.  Furthermore, the amount of dust extinction
in the integrated stellar continuum has been observed to be less than that 
from star-forming regions, as probed through recombination line ratios 
in local starbursting galaxies \citep{Calzetti00}.  
It is important to understand the amount of
dust extinction in galaxies, as extra extinction will result in an underestimate
of star formation rates derived from the H$\alpha$ line fluxes relative to those derived from the
bolometric rest-frame ultraviolet luminosity.  Studies of
the dust properties in distant star-forming galaxies are not conclusive:
while \citet{Erb06} and \citet{Reddy10} find no evidence for extra extinction
towards star-forming regions, numerous other studies do measure varying degrees of
extra extinction \citep{ForsterSchreiber09, Yoshikawa10, Mancini11, Wuyts13, Kashino13, Price13}.

Global measurements of the Balmer decrement have been made for a few individual distant galaxies at $z>1.5$ 
\citep{Teplitz00, vanDokkum05, Hainline09}, or through stacking analyses of larger samples of up to a few 
hundred galaxies at $z>1.5$ \citep[e.g.,][]{Yoshikawa10, Kashino13}. 
As these measurements were made using ground-based near-infrared (NIR) spectrographs, sky lines can severely contaminate 
the line fluxes.  Ideally, one would study the nature of dust in distant galaxies using 
space-based spectroscopy \citep[e.g.,][]{Price13}.
Here, for a single galaxy, we are able to measure the Balmer decrement through the ratio of the 
H$\gamma$ and H$\beta$ line fluxes from the space-based HST/WFC3 NIR grism spectroscopy,
spatially-resolving the dust extinction for six \emph{individual} star-forming regions for the first time 
in a distant galaxy.

\begin{figure}[t]
\leavevmode
\centering
\includegraphics[width=0.98\linewidth]{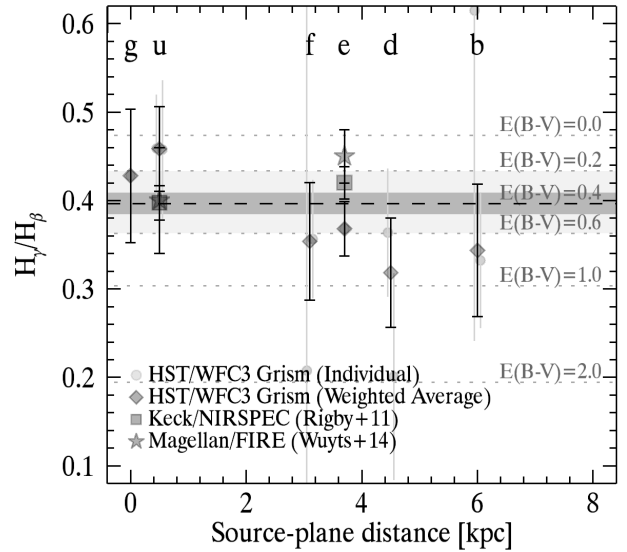}
\caption{Resolved H${\gamma}$/H${\beta}$ line ratios in 6 star-forming clumps show no significant variation
across the galaxy.  Where multiple grism
measurements for the same clump are available (light grey circles), the weighted average is shown (black diamonds).  
Keck/NIRSPEC \citep[][squares]{Rigby11} and Magellan/FIRE \citep[][stars]{EWuyts14} measurements are also included. The
weighted average of all data is $E(B-V)_{gas}=0.40\pm0.07$ (dashed line) with a standard deviation of 0.2.  }
\label{fig:HbHg}
\end{figure}

Figure~\ref{fig:HbHg} shows the H${\gamma}$/H${\beta}$ line ratios for six star-forming clumps, 
as a function of their source-plane distance, as derived in \citet{EWuyts14}.  We exclude clump \emph{c} here, as the 
H$\gamma$ emission line flux is not well constrained.  When defining this 
source-plane distance, clump \emph{g} is assumed to be the center of the galaxy.  This choice is
justified by the fact that clump \emph{g} is the only star-forming region to show up in the
stellar mass surface density maps presented in \citet{EWuyts14}.
We assume an absorption correction of 5$\mathrm{\AA}$ to the equivalent widths of both Balmer lines, following
predictions from low-metallicity (Z=0.008) simple stellar populations in \citet{Groves12}.  
For multiple images of 
the same star-forming region, the individual measurements are shown as light grey circles with the weighted
average indicated with a diamond.  We include additional
Balmer decrement measurements from ground-based Keck/NIRSPEC \citep[][squares]{Rigby11} and Magellan
Folded-port InfraRed Echellette (FIRE) data \citep[][stars]{EWuyts14}.
Although the grism data alone shows a trend for higher dust extinction with increasing distance from the established
stellar population in clump \emph{g}, with a best-fit linear slope of $-0.02\pm0.01$, including the ground-based data yields
a best-fit slope of $0.00\pm0.01$, consistent with constant extinction across all star-forming regions in the galaxy.  
Down-weighting the ground-based measurements to account for systematic uncertainties in the telluric contamination
does not change this result.
We find the weighted average of all of the data to be $E(B-V)_{gas}=0.40\pm0.07$ (dashed
line in Figure~\ref{fig:HbHg}), with a standard deviation of 0.2. Due to the large error bars in the HST/WFC3 
measurements, we do not detect significant variations in extinction between the different clumps and therefore adopt the average
value for future dust corrections.

In a separate analysis, \citet{EWuyts14} measure an average stellar extinction of $E(B-V)_{star}=0.28\pm0.04$ 
from spectral energy distribution (SED) modeling.
The linear ratio of the average extinction in HII regions as probed by the emission line ratios presented here,
relative to the extinction of the stellar light from the SED modeling is $1.4\pm0.3$.  This indicates
extra extinction towards HII regions, consistent with measurements by \citet{Price13} and
\citet{Wuyts13}.

\section{The Star Formation Sequence}
\label{sec:sfr}

Galaxies are observed to show a correlation between their star formation rate 
(SFR) and stellar mass (M$_{\star}$) from the present
day until the earliest observed epoch, less than a billion years after the Big 
Bang \citep[e.g.,][]{Brinchmann04, Noeske07a, Elbaz07,
Daddi07, Pannella09, Magdis10, Gonzalez10, Whitaker12b, KGuo13}.  
The characteristics of this relatively tight correlation are linked with the 
physical processes regulating galaxy formation and evolution itself \citep{Dutton10, Leitner12}.  

\begin{figure}[t]
\leavevmode
\centering
\includegraphics[width=0.98\linewidth]{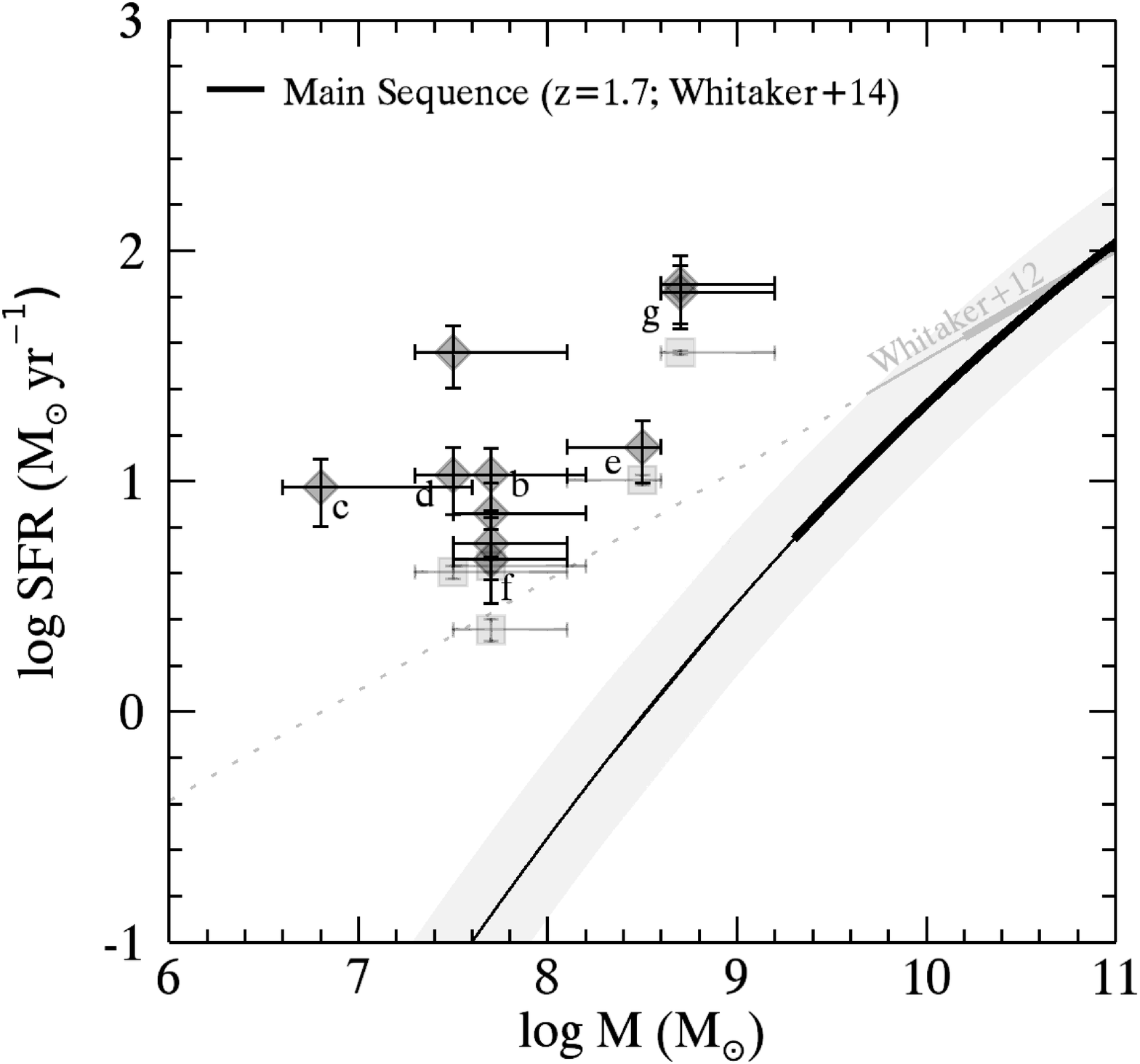}
\caption{Star formation rate estimates from demagnified HST/WFC3 grism H$\beta$ line fluxes (diamonds) and 
H$\alpha$ line fluxes 
from ground-based measurements \citep[squares,][]{EWuyts14} for individually resolved
star-forming clumps in RCS0327.  We correct for an average measured extinction of $\mathrm{E(B-V)_{gas}}=0.4$,
where the HST/WFC3 error bars correspond to the error in the emission line fluxes added in quadrature to uncertainty in 
the mean extinction (0.07 mag, black) and 1$\sigma$ spread in the individual extinction measurements (0.2 mag, grey).
The ground-based H$\alpha$ SFRs are 0.2--0.4 dex lower due to aperture correction uncertainties.
The SFRs of the
individual clumps lie ubiquitously above the star formation sequence at $z=1.7$ from Whitaker et al. in prep
(black) and \citet{Whitaker12b} (silver).  Thick lines indicate mass-complete samples, thin
lines are mass-incomplete, and the dotted line is an extrapolation.}
\label{fig:sfr}
\end{figure}

Star formation within galaxies
is believed to be regulated by the balance between gas accretion and feedback preventing this gas from cooling,
where fluctuations in the gas accretion rate and star formation efficiency are thought to be encapsulated in the intrinsic scatter of
the star formation sequence.
This scatter has been observed to be correlated with gas mass fractions \citep{Magdis12}, environment 
\citep{Patel11}, morphology \citep{Wuyts11b}, and galaxy merger and interaction processes \citep{Lotz08, Jogee09}.  This last correlation 
is of particular interest for RCS0327, an ongoing merger \citep{EWuyts14}.  
\citet{Jogee09} demonstrate that interacting galaxies out to $z=0.8$ are observed to have the highest
SFRs for a given M$_{\star}$, consistent with the predicted star formation histories from 
the hydrodynamical merger simulations of \citet{Lotz08}.  

SFR estimates from the HST/WFC3 grism H$\beta$ line fluxes of the individual star-forming clumps of RCS0327 are 
presented in Figure~\ref{fig:sfr} (diamonds).  The H$\beta$ fluxes are demagnified by 
dividing by the median magnification from the lensing map within the extraction region for each spectrum.
We correct the H$\beta$ line fluxes for the average measured 
extinction of $\mathrm{E(B-V)_{gas}}=0.4\pm0.07$ and assume an intrinsic ratio of (H$\alpha$/H$\beta$)$_{\mathrm{int}}$=2.86
\citep{OsterbrockFerland} to convert to H$\alpha$.
We further correct these line fluxes for the expected contribution from outflows, ranging from 30--55\%, 
as measured for the H$\alpha$ broad and narrow-line profiles from
OSIRIS integral field spectroscopy presented in Table 3 of \citet{EWuyts14}.  
We also include SFRs derived from the OSIRIS demagnified H$\alpha$ luminosities where available \citep[squares,][]{EWuyts14},
assuming the same dust and outflow corrections.
The H$\alpha$ luminosities are converted to SFR 
following the calibration in \citet{Kennicutt98}, plus a factor of 1.7 correction to convert from 
\citet{Salpeter} to a \citet{Chabrier} IMF.  In Figure~\ref{fig:sfr}, we see that the ground-based SFRs
are systematically lower than the space-based SFRs.  This difference likely results from aperture correction 
uncertainties in the ground-based data.  The OSIRIS H$\alpha$ fluxes are measured in an aperture derived from a 2D 
gaussian fit to the H$\alpha$ image.  This aperture effectively probes the ionized core of the HII region and does 
not capture the diffuse outer regions.  Also, the AO corrections will spread out the light.  
Together, these effects result in systematically lower ground-based derived SFRs.
Stellar masses are derived from SED modeling of the individual clumps, as described in \citet{EWuyts14}.

In the case where there are multiply lensed images of the same physical region,
we find that the measured H$\beta$ line fluxes and inferred SFRs are consistent within the error bars.  
There is one notable exception; clump \emph{d} has a widely varying SFR that is not understood.  
Although the observed H$\beta$ fluxes in the 
different spectra have similar magnitudes, the magnification is significantly smaller in the counter-arc.  Adopting the 
individual extinction measurements does not bring these measurements into agreement.

In \citet{Whitaker12b}, we measured the SFRs out to $z=2.5$ with the accurate rest-frame colors and photometric redshifts
from the NEWFIRM Medium-Band Survey \citep[NMBS,][]{Whitaker11} by adding the bolometric
rest-frame ultraviolet luminosity to the bolometric infrared luminosity, as calibrated from Spitzer 
Multiband Imaging Photometer (MIPS) 24$\mu$m imaging.
A measurement with mass-completeness limits extending to signficantly lower stellar masses from 
Cosmic Assembly Near-infrared Deep Extragalactic Legacy Survey (CANDELS) and 3D-HST data
is presented in Whitaker et al. (in prep)\footnote{Through a stacking 
analysis of the 24$\mu$m imaging, Whitaker et al. (2014)
circumvent the SFR completeness limits in NMBS.  The trade off
with this technique is that while the average $\log\mathrm{SFR}-\log\mathrm{M_{\star}}$  
relation can be measured robustly,
one cannot formally measure the intrinsic scatter at low stellar masses.}.
In Figure~\ref{fig:sfr}, we compare the resolved SFRs for the star-forming clumps to the 
average observed relation at $z=1.7$ from Whitaker et al. in prep
(black) and \citet{Whitaker12b} (silver).  Thick lines indicate mass-complete samples, thin lines are mass-incomplete and the
dotted line is an extrapolation.
We note the danger in extrapolating the observed star formation sequence far below the mass-completeness limits:
as demonstrated in Figure~\ref{fig:sfr}, the normalization is over-estimated by almost 2 dex and the slope is under-estimated
at stellar masses of 10$^{8}$ M$_{\odot}$ when extrapolating the NMBS data at $z=1.7$, as 
compared to the deeper 3D-HST/CANDELS analysis of Whitaker et al. in prep 
(black)\footnote{The NMBS sample of \citet{Whitaker12b} is incomplete in M$_{\star}$ and SFR 
at $\log\mathrm{M_{\star}}\sim10$, resulting in a $\sim0.1$ dex offset (thin grey line) largely
driven by the limited depth of the MIPS/24$\mu$m imaging. Only galaxies with the 
highest sSFRs will be detected and the average relation will consequently be overestimated.}.

We see that all star-forming regions lie $>$1--2 dex 
above the average star formation sequence observed in a similar mass range by Whitaker et al. (in prep).
We cannot compare to \citet{Whitaker12b}, as there is no data in this parameter space to justify 
the extrapolated slope. 
The shaded region in Figure~\ref{fig:sfr}
shows the measured intrinsic scatter of 0.25 dex in the $\log\mathrm{SFR}-\log\mathrm{M_{\star}}$  
relation from \citet{Whitaker12b}.
Through a compilation of 25 studies
in the literature, \citet{Speagle14} find a roughly constant intrinsic scatter
for the $\log\mathrm{SFR}-\log\mathrm{M_{\star}}$ relation both with redshift and
stellar mass.  However, the intrinsic scatter may be larger at low stellar masses; dwarf galaxies show diverse
\citep{Weisz11} and stochastic \citep{Bauer13} star formation histories.  
Although the scatter at the low mass end is not well understood, the intrinsic scatter in $\log\mathrm{SFR}$
would need to increase by a factor of $>4$ for the star-forming regions of RCS0327 to
no longer be considered outliers.

The average specific SFR ($\mathrm{sSFR}\equiv\mathrm{SFR/M}_{\star}$) 
from the HST/WFC3 measurements is 10$^{-6.7}$ yr$^{-1}$ with a scatter of 0.5 dex.
The relatively small observed intrinsic scatter along the main star-forming sequence of 0.25 
dex measured by \citet{Whitaker12b} implies that the individual star-forming clumps all lie 0.75--2.75 dex
above the $1\sigma$ envelope. 
The average sSFR and scatter are unchanged when correcting with the individual $\mathrm{E(B-V)_{gas}}$ 
measurements presented in Figure~\ref{fig:HbHg}, instead of the average value.
These observations confirm previous results that galaxy mergers have enhanced SFRs relative to the ``normal'' star formation sequence.
Furthermore, the similar individual measured sSFRs imply that a global estimate for this galaxy would be generally
representative of the star formation activity on sub-galactic scales.
Given high SFRs and the relatively low inferred dust extinction across the
entire galaxy presented in Section~\ref{sec:dust}, hydrodynamical merger simulations would suggest 
that RCS0327 is in the mid- to final merger stages \citep{Lotz08}.

\section{High Redshift Ionization Conditions}
\label{sec:bpt}

\subsection{AGN Search}
\label{sec:agn}

We searched for the presence of an active galactic nucleus (AGN) in RCS0327 using three different methods. 
First, we examined the position of individual knots on the \citet{Baldwin81} diagnostic diagram 
(henceforth the BPT diagram) and find that the line ratios of
all star-forming regions safely fall within the HII-origin parameter space (see Section~\ref{sec:bptdiagram}).
Second, we obtained a 60~ks Chandra image through the Cycle 12 
Guest Observer Program (Observation ID 12960, PI:Rigby) and 
searched for point sources near the nucleus.
No point sources coincident with knots \emph{e} and \emph{g} were seen in the Chandra data, 
setting an upper limit on the rest-frame 2--10~keV
luminosity of an AGN at $1\times10^{43}$~erg~s$^{-1}$ (knot \emph{e}) and $7\times10^{43}$~erg~s$^{-1}$ (knot \emph{g}),
if we assume Galactic absorption.  If we assume Compton-thick intrinsic absorption ($N_{H}=2\times10^{24}$ cm$^{-2}$),
then the upper limits are $\sim1\times10^{45}$~erg~s$^{-1}$ (knot \emph{e}) and $\sim7\times10^{45}$~erg~s$^{-1}$ (knot \emph{g}).
Although we cannot rule out the possibility of a highly obscured luminous AGN, a 10$^{45}$~erg~s$^{-1}$ AGN is more than 100
times rarer than a 10$^{43}$~erg~s$^{-1}$ AGN at $z\sim1.7$ \citep[e.g., ][]{Aird10}.  From the X-ray 
upper limits alone, we cannot rule out the presence of a moderate luminosity AGN.
Third, we difference two HST F814W images of 2100~s depth taken three months apart (Program 12371 PI:Rigby, 
and Program 12267, PI:Rigby).  Subtraction of the two epochs of HST images reveals no significant variability.
This AGN search will be presented in full in a future paper.

\subsection{Dissecting the BPT Diagram}
\label{sec:bptdiagram}

A standard method of separating line emission originating from HII regions from that of gas photoionized 
by a harder radiation field
is using the $\mathrm{[N\Rmnum{2}]}$/H${\alpha}$ versus $\mathrm{[O\Rmnum{3}]}$/H${\beta}$ line ratios.
The BPT diagnostic diagram was first proposed by \citet{Baldwin81}, and later
 refined by \citet{Veilleux87} and \citet{Kewley01}.  
A hard ionization field can either be produced from shock excitation or the accretion
disk of an AGN.  In the absense of this hard radiation field, star-forming
galaxies form a tight sequence on the BPT diagram, often referred to as the star-forming or HII
abundance sequence \citep{Dopita86}.  The location of this
abundance sequence probes the metallicity, stellar ionizing radiation field and physical conditions of the
interstellar medium surrounding star-forming regions.    As detailed by \citet{Kewley13}, a harder
ionizing radiation field and/or a larger electron density moves galaxies above the normal star-forming
abundance sequence, whereas a larger ionization parameter can either raise or lower the line ratios,
depending on the metallicity.  

\citet{Kewley13} provide theoretical predictions for the lower limit abundance sequence, as derived from 
``normal'' star-forming conditions which assume the same shape extreme ultraviolet radiation field, electron densities and 
relationship between metallicity and ionization parameter as local star-forming galaxies.  In this case,
chemical evolution will move star-forming galaxies toward smaller $\mathrm{[O\Rmnum{3}]}$/H${\beta}$ and
larger $\mathrm{[N\Rmnum{2}]}$/H${\alpha}$ line ratios with cosmic time.  The upper limit abundance sequence
proposed by \citet{Kewley13} propagates extreme star-forming conditions resulting from either a hard radiation 
field or a combination of a larger ionization parameter ($\log U > -2.9$) and larger electron 
densities ($n_{e}\sim1000$ cm$^{-3}$) than local galaxies for 
metallicities below $\log\mathrm{(O/H)}+12\sim8.8$.  This upper limit star-forming abundance sequence evolves
with redshift with an uncertainty of approximately 0.1 dex on the BPT diagram at a given redshift.

\begin{figure}[t]
\leavevmode
\centering
\includegraphics[width=0.95\linewidth]{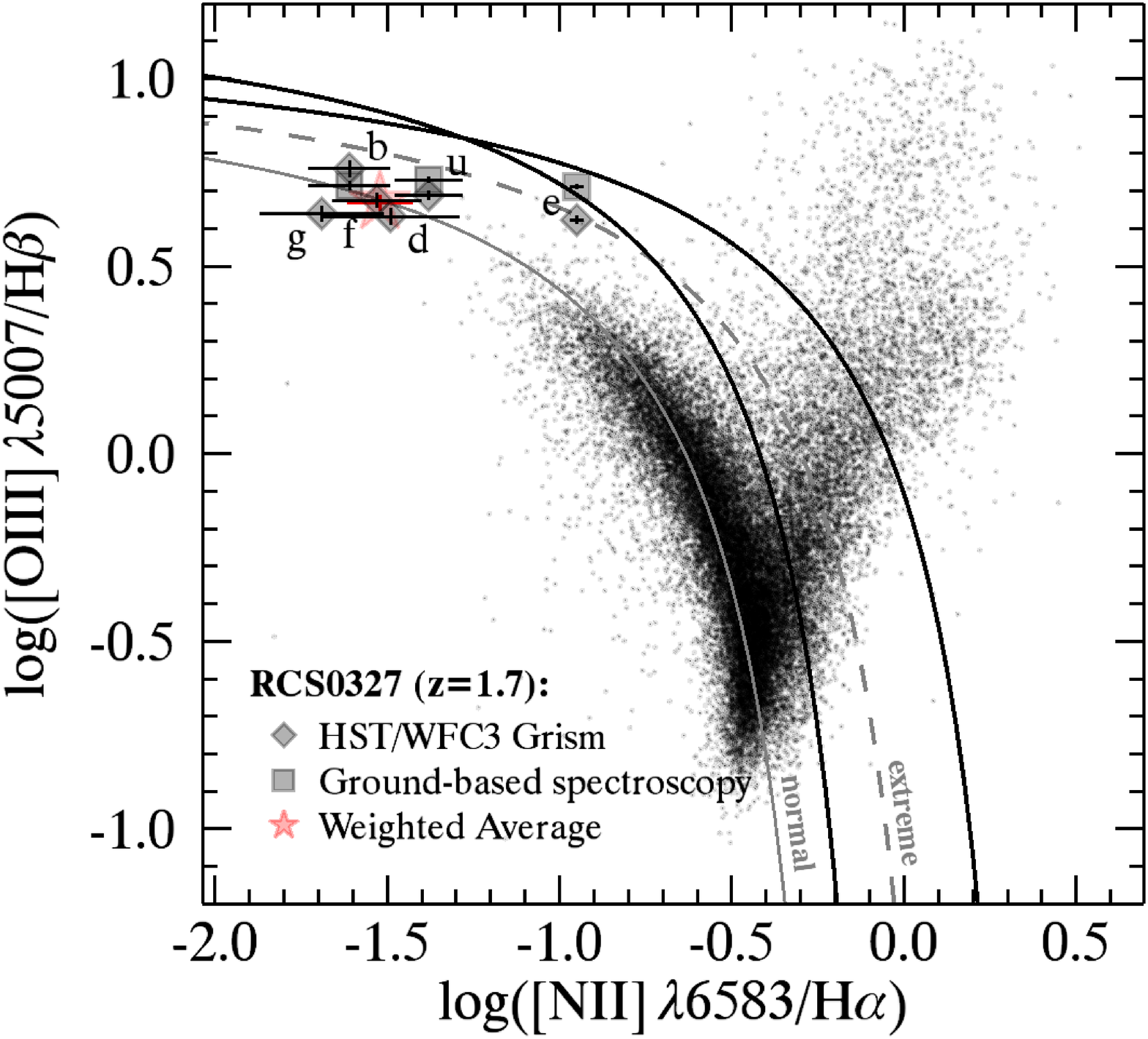}
\caption{BPT diagram.  The local relation is defined by the SDSS DR7 spectroscopic sample \citep[][black dots]{Abazajian09},
with \citet{Kauffmann03b} and \citet{Kewley01} empirical and theoretical divisions between emission from
HII regions and AGN (black).  The resolved star-forming regions of RCS0327 from this work are indicated with
diamonds, where clump \emph{e} is offset from
the other regions. We include additional Keck/NIRSPEC observations of clumps \emph{e} and \emph{u} \citep{Rigby11} and 
Magellan/FIRE observations of
clump \emph{b} \citep{EWuyts14} (square symbols).  The red star indicates the average weighted by the
demagnified F160W flux for each star-forming region in
RCS0327. The local stellar abundance sequence
representing ``normal'' physical conditions (solid grey) coincides with the majority of the star-forming clumps in RCS0327,
whereas clump \emph{e} lies along the predicted evolution for the stellar abundance sequence for extreme physical conditions
at $z=1.7$ (dashed grey) from \citet{Kewley13}.}
\label{fig:bpt}
\end{figure}

High redshift observations seem to indicate that the bulk of galaxies
lie above the local abundance sequence \citep[e.g.,][]{ForsterSchreiber09,
Yabe12, Yabe14, SNewman14, Zahid13b, Cullen13}.  However, it is unclear whether high redshift star-forming galaxies
have more extreme in situ star-forming conditions or have a contribution to the line emission from AGN and/or shock excitation.
The majority of $z\sim2$ studies are limited to massive, high metallicity galaxies \citep[e.g.,][]{ForsterSchreiber09,
Yabe12} that dominate 
the stellar mass density budget of the universe, whereas
we present data for a lower metallicity and stellar mass galaxy that 
is representative of the regime where galaxies have the largest 
number densities \citep{Marchesini09}.

In this work, we take advantage of both the magnification from the gravitational lens and the high spatial resolution
provided by the WFC3 grism to dissect the origin of line emission in six resolved star-forming clumps of RCS0327.  
We exclude clump \emph{c} in this analysis, as $\mathrm{[N\Rmnum{2}]}$ and H${\alpha}$ line fluxes are not available.
Figure~\ref{fig:bpt} shows the BPT diagram for RCS0327 relative to the local abundance sequence and empirically and theoretically
derived 
relations.  As detailed in Section~\ref{sec:agn}, we can rule out the presence of an unobscured high luminosity AGN in RCS0327.

Next we explore the reasons why clump \emph{e} is offset in the BPT diagram.
We find little evidence for an extreme ionization parameter in clump \emph{e} relative to the other star-forming regions. 
The $\mathrm{[O\Rmnum{3}]}$/H${\beta}$ line ratios agree within 0.2 dex for all star-forming clumps.
However, clump \emph{e} does have a higher electron density of $600\pm100$ cm$^{-3}$, as measured from the [OII] 3726,3729 
doublet ratio by \citet{EWuyts14} that is independent of the lensing magnification.
\citet{EWuyts14} measure outflow wind velocities of 145 to 315 km/s in clumps \emph{b, d, e} and \emph{f}.
Clump \emph{e} is observed to have the highest outflow velocity (315 km/s)
and relative contribution of H$\alpha$ flux in the broad 
component (54\%). 

Contribution from shock excitation may explain the peculiar behavior of clump \emph{e} on 
the BPT diagram.
Radiative shocks produce a hard ionizing radiation field
that can also contaminate the measured line ratios.
Slow shocks associated with galactic-scale winds have been observed locally 
\citep{Rich10,Rich12} and at $z=1$ \citep{Yuan12}.  
These shocked stellar winds can change the geometrical distribution of the gas
around the ionizing source, and hence the stellar
ionizing radiation field.  Stellar winds will preferentially clear ionized gas from the interior of an H~II region,
lowering the effective ionization parameter \citep{Yeh12}.  
The higher measured outflow in clump \emph{e} will move
it down and to the right of the global average for RCS0327 in the BPT diagram.  When combining this with the higher
electron density, which will move clump \emph{e} up and to the right in the BPT diagram, we find
that clump \emph{e} effectively moves to the right.  We suspect that the combination of shocks 
and a slightly higher electron density could result in
clump \emph{e} lying on the upper limit ``extreme'' abundance sequence predicted for $z=1.7$ in Figure~\ref{fig:bpt}.

An alternative explanation of why clump \emph{e} is offset in the BPT diagram is that it 
hosts an AGN.  Using spatially-resolved measurements from the SINS survey \citep{ForsterSchreiber09}, \citet{SNewman14} find
that the inner regions of $z\sim2$ galaxies are offset to
higher excitation relative to their outer regions.  
Active nuclei could explain the higher excitation found in the central regions
of galaxies by \citet{SNewman14}.
Without the added spatial resolution and magnification
from gravitational lensing, it is difficult to distinguish the origin of this offset given the
large range of possible physical mechanisms listed above.  Here, we do not find such a trend, with five
out of six star-forming regions following the local abundance sequence (including the ``central'' clump \emph{g}).  
Even if we instead assume that clump \emph{e} is the center of the galaxy, 
contrary to the spatially-resolved stellar mass analysis,
the evidence still shows that clump \emph{e} does not appear to host an AGN.

\subsection{He~I/H$\beta$ Line Ratio Diagnostics} 
\label{sec:HeI}

\begin{figure*}[t]
\leavevmode
\centering
\includegraphics[width=0.75\linewidth]{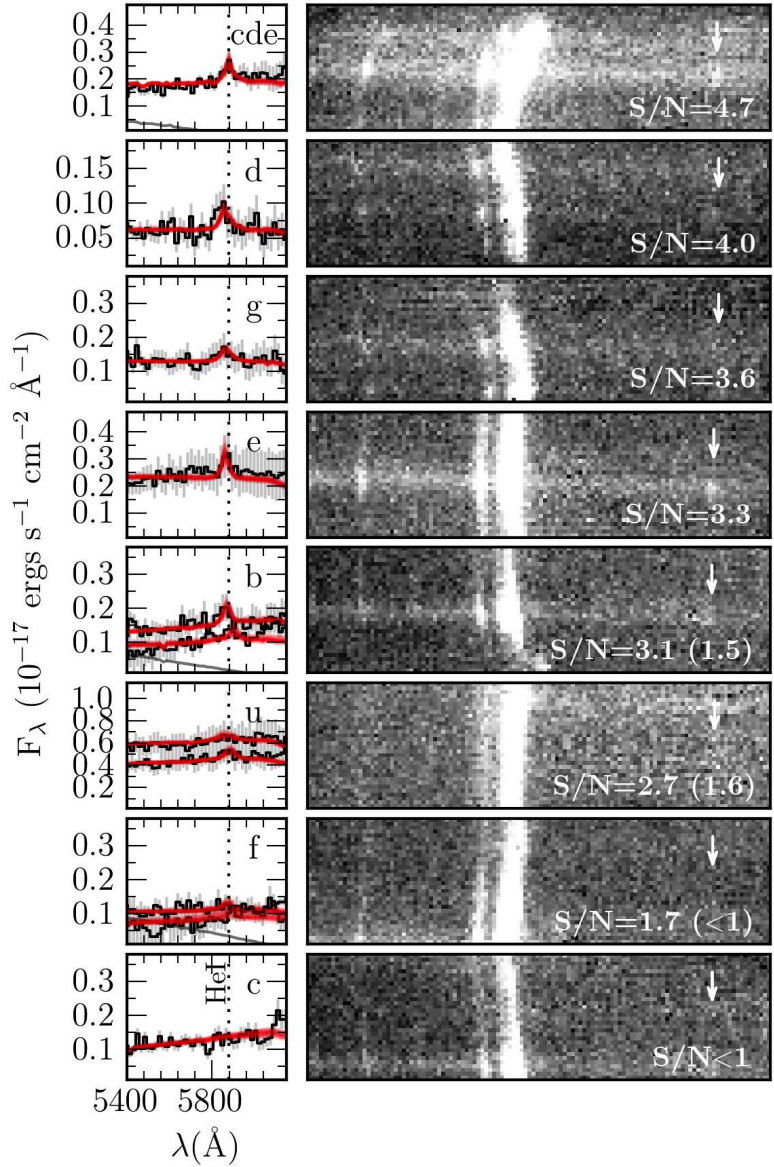}
\caption{One dimensional grism spectra for individual resolved star-forming regions around the He~I $\lambda5876$ line (left), 
with the best-fit convolved model spectra from a Markov chain Monte Carlo analysis (red) and contamination (grey).  
The contamination is only significant in clump \emph{b} and \emph{f}.  The spectra are rank ordered by S/N ratio,
and the values in parentheses indicates the measured S/N for the second observation of the same physical region.
The two dimensional grism spectra for the full wavelength range (right), with an arrow indicating the location of the He~I line,
demonstrate the robust detection of the faint He~I lines in most cases (although clump \emph{f} is an upper limit, clump
\emph{u} has significant morphological broadening, and there is no formal detection of the faintest clump \emph{c}).}
\label{fig:HeIspectra}
\end{figure*}

\begin{figure}[t]
\leavevmode
\centering
\includegraphics[width=0.98\linewidth]{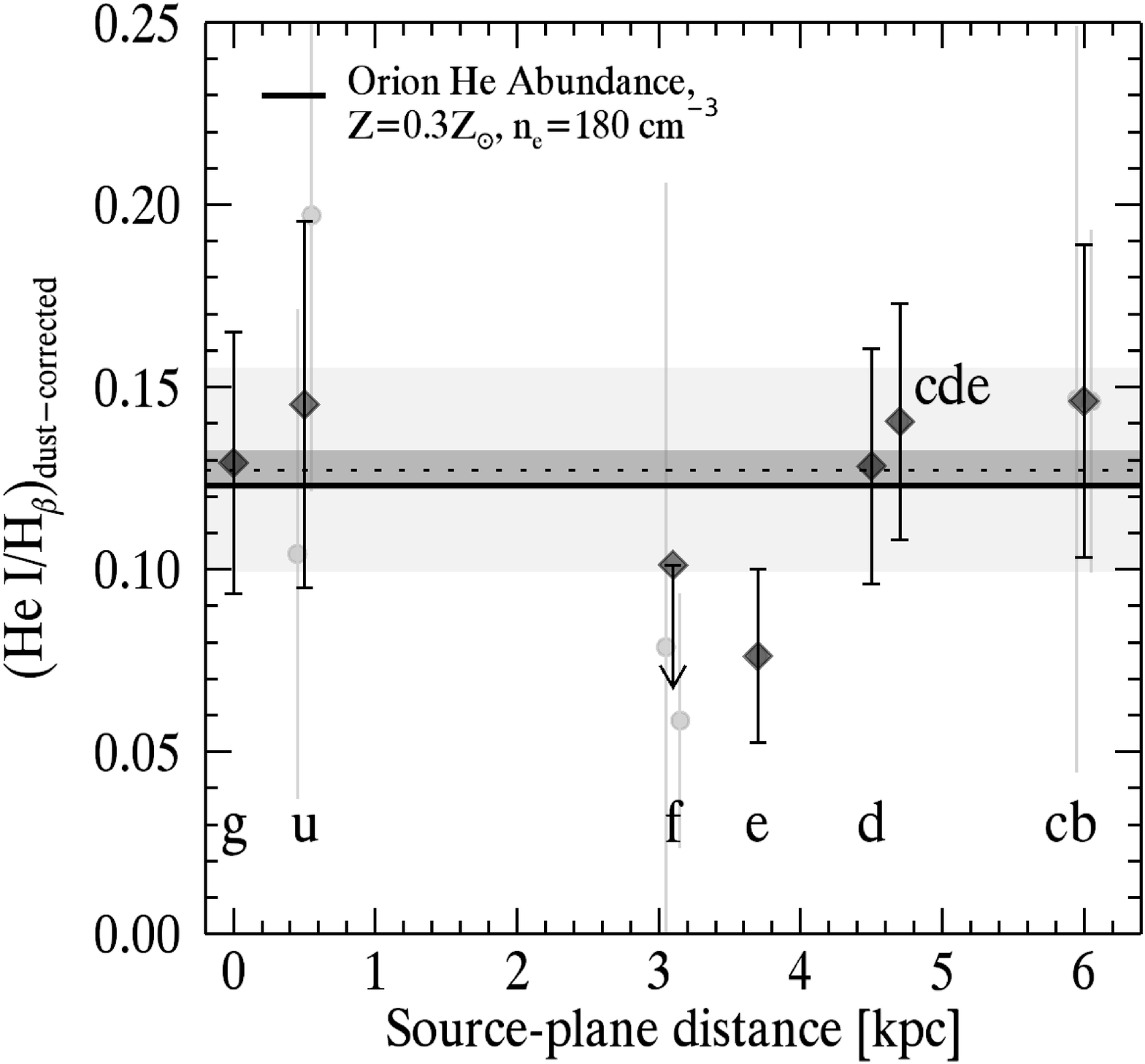} 
\caption{The He I $\lambda5876$ to H$\beta$ line ratios in five resolved star-forming regions of a
lensed galaxy merger at $z=1.7$ are consistent with the saturated value, implying that their ionizing spectra
are similar to that of a \citet{Salpeter} present-day mass function extending above 60 M$_{\odot}$.
In the case of multiple grism measurements for the same clump (light grey circles), the weighted average
is shown (black diamonds).  The average ratio, weighted by the demagnified F160W flux for each
star-forming region, the error in the average ratio, and the $1\sigma$ scatter is indicated by
the dotted line, dark and light shaded region, respectively. The solid black line is the predicted
ratio from Cloudy models, assuming the Orion He abundance (equivalent to the Solar He abundance),
30\% Solar metallicity, and an electron density of 180 cm$^{-3}$.  The He I $\lambda5876$/H$\beta$
ratio scales linearly with He abundance.}
\label{fig:HbHeI}
\end{figure}

We detect the He~I $\lambda5876$~\AA\ emission line in several knots of the WFC3 grism data of
RCS0327.  As the He~I $\lambda5876$~\AA\ emission line has not been used previously, to our knowledge, to 
study distant galaxies, we now explore the origin of this emission.  We then examine the
strength of the line in the bright knots of RCS0327, compare to
theoretical predictions from the Cloudy plasma code, and comment upon
the massive star content of these star-forming regions.

Because it is a helium recombination line,  He~I~$\lambda5876$ measures the
nebular ionization rate, and thus, probes how many extremely massive 
stars are present in a star-forming region.  The ratio of a Helium
recombination line to a Hydrogen recombination line has long been recognized as a
diagnostic that is sensitive to the relative sizes of the He~II and
H~II Stromgren spheres \citep{Bashkin75}.  As \citet{OsterbrockFerland} explain,
there are two limiting cases.  In the case of a soft ionizing spectrum, there are far more photons
capable of ionizing H than photons capable of ionizing He, resulting
in a He~II Stromgren sphere much smaller than the H~II
Stromgren sphere.  In the case of a hard ionizing spectrum, the ionizing spectrum contains 
so many photons with energies exceeding the first ionization potential
of He (24.6 eV), that these photons dominate the ionization of both H
and He.  In this hard case, the H~II and He~II
Stromgren spheres overlap, and there is a single region in which
both H and He are ionized.  

Simple models \citep[][Figures 2.4 and 2.5]{OsterbrockFerland} show that
this hard case occurs, i.e. the Stromgren spheres overlap,  when the
stellar effective temperature is at least 40,000~K.  This is roughly equivalent to an O7 V star 
\citep[Table 2.3 of][]{OsterbrockFerland}, with a mass of $\sim35$~M$_{\odot}$~\citep{Gies02, Niemela94}.  
Once the Stromgren spheres
overlap, a He~I to H~I recombination line ratio saturates at its
maximum value; successively harder spectra do not change the ratio.
The value at which the line ratio saturates is sensitive to the He
abundance, and is insensitive
to density, ionization parameter, electron temperature, or other
nuisance variables \citep{Vanzi96,ForsterSchreiber01}.

Although such simple models are appropriate for the simplest Galactic H~II
regions where a single star dominates the ionizing continuum \citep{Oey00,Kennicutt00}, it is
not appropriate to characterize the  ionizing spectrum with an
effective temperature when the ionizing source is not one hot star,
but many hot stars, as in the case of large H~II regions or super star clusters.
However, a simple rule-of-thumb emerges:  a region
that is forming stars continuously with a \citet{Salpeter}-slope IMF extending past
40--60 M$_{\odot}$ will have a sufficiently hard enough spectrum that the
Stromgren spheres overlap, and a He~I to H~I
line ratio will saturate \citep{Rigby04}.  

\subsubsection{He~I/H$\beta$ Line Measurements}

The 2D spectra in the right panel of Figure~\ref{fig:HeIspectra} show a 
clear He~I $\lambda5876$ detection in \emph{four} of the seven
star-forming knots at $\mathrm{S/N}>3$.  Among the cases that lack an obvious detection, the He~I line has large morphological
broadening in knot \emph{u}, we report a $1\sigma$ upper limit for the marginal detection in knot \emph{f},
and there is no formal detection in the faintest knot \emph{c}.
The 1D spectra for all grism extractions of the star-forming regions 
are shown in the left panels of Figure~\ref{fig:HeIspectra}, ranked by S/N.

By combining the He~I $\lambda5876$ measurements with the well-detected H$\beta$ line fluxes,
we make the unique measurement of the number of Helium to Hydrogen ionizing photons
for individual star-forming regions within a merging galaxy at $z=1.7$.  These
ratios have been corrected for the average line ratio of $\mathrm{H}\gamma/\mathrm{H}\beta=0.40\pm0.04$
measured in Section~\ref{sec:dust}, converted into an
average dust extinction within the galaxy assuming the \citet{Calzetti00} law.  We note that
although correcting by the individual $H\gamma/H\beta$ measurements presented in Figure~\ref{fig:HbHg} does not
significantly change the results, we chose to adopt the average correction.

\begin{figure*}[t]
\leavevmode
\centering
\includegraphics[width=0.85\linewidth]{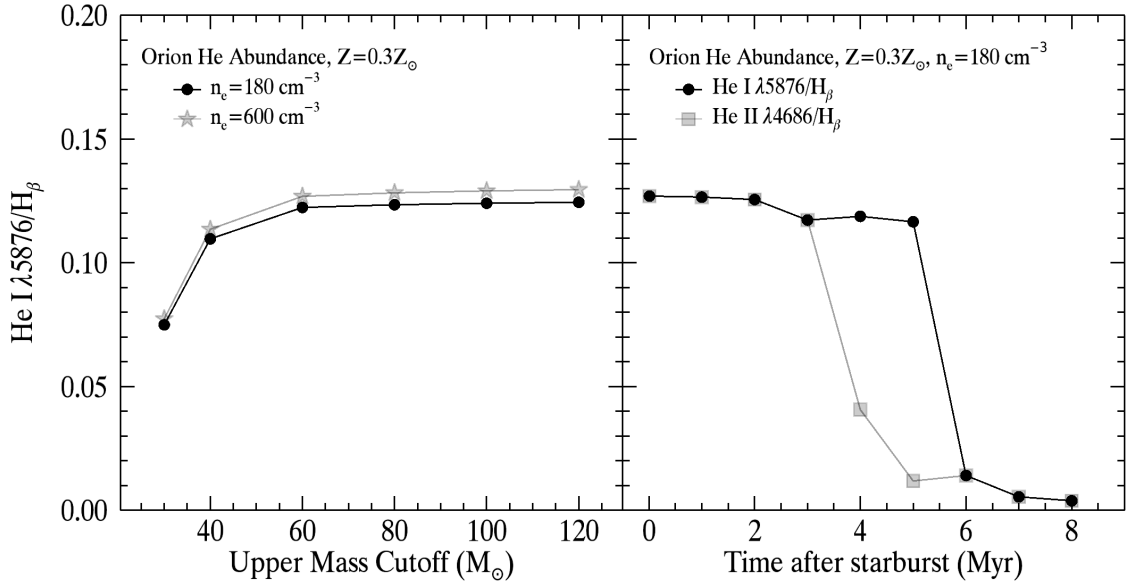}
\caption{(Left) Theoretical He I $\lambda5876$/H$\beta$ line ratios from Cloudy models, assuming 
continuous star formation for an Orion He abundance pattern, 30\% Z$_{\odot}$ and two electron densities (180 and 600 cm$^{-3}$).
The line ratio saturates for a \citet{Salpeter} IMF with an upper mass cutoff greater than or equal
to 60 M$_{\odot}$.
While the line ratio is insensitive to electron density and gas-phase metallicity, there is a
linear relationship with He abundance.  Higher He abundances results in higher 
He I $\lambda5876$/H$\beta$ ratios. (Right)  Evolution of theoretical He I $\lambda5876$/H$\beta$ and
He II $\lambda4686$/H$\beta$ line ratios as an instantaneous starburst ages. The He II $\lambda4686$/H$\beta$ line ratio drops steeply first, followed by a drop in the He I $\lambda5876$/H$\beta$ 3 Myr later. }
\label{fig:HeI_masscutoff}
\end{figure*}

In Figure~\ref{fig:HbHeI}, we see that the majority of the clumps have a similarly high He~I $\lambda5876$ to 
H$\beta$ ratio, with a weighted mean value of $0.135\pm0.005$ and a standard deviation of 0.028.  Both measurements
include the uncertainty in the mean dust extinction correction.  
The ratio measured for the high S/N spectra of 
knot \emph{e} is $2\sigma$ below the average
value, as weighted by the demagnified F160W flux within each extraction region rather than the individual measurement errors.  
Similarly, we present the $1\sigma$ upper limit for knot \emph{f}.

In addition to He I $\lambda5876$, we also formally detect He II $\lambda4686$ in knot \emph{g} with a line flux 
measurement of $1.64\pm0.90\times10^{-17}$ erg s$^{-1}$ cm$^{-2}$.  Together with our H$\beta$ line fluxes, we find
a He II $\lambda4686$/H$\beta$ line ratio of $0.12\pm0.07$ in knot \emph{g}.  We also 
constrain the He II $\lambda4686$/H$\beta$ line ratio for knot \emph{e}, finding a value of $0.03\pm0.02$.  
The error bars for the remainder of the knots are too large to constrain the He II $\lambda4686$/H$\beta$
line ratio.

\subsubsection{Models}

To analyze the He~I~$\lambda5876$ and He~II~$\lambda4686$ to H$\beta$ line ratios we observe in RCS0327,
we generate theoretical spectra of star-forming regions using the stellar
synthesis code Starburst~99 \citep{Starburst99, Vazquez05, Leitherer10} and the plasma code
Cloudy \citep{Ferland13}.  We generate ionizing spectra using Starburst~99 (version 6.0.4), 
assuming continuous star formation with a metallicity $40\%$ of solar\footnote{Starburst99
has only a set number of stellar spectral templates; 40$\%$ of solar
is the closest match to the measured $30\pm4\%$ metallicity measured by \citet{EWuyts14} 
for RCS0327.}, and a \citet{Salpeter} IMF high-mass slope\footnote{The slope of the IMF at the massive
end is the same for the \citet{Chabrier} IMF, adopted throughout the rest of this paper.
The ionizing spectrum originates solely from the high mass end where
the \citet{Chabrier} and \citet{Salpeter} IMFs will produce identical results.}.
We feed the ionizing
spectrum from Starburst99 into Cloudy \citep[version 13.02,][]{Ferland13}
assuming 30$\%$ solar metallicity, a hydrogen density of
either 180~cm$^{-3}$  or 600~cm$^{-3}$ (as measured by \citet{EWuyts14}, using
data from \citet{Rigby11} for knots \emph{u} and \emph{e}, respectively), and an ionization
parameter of $\log U = -2.84$, which is the value measured by \citet{Rigby11}
scaled for the average extinction we measure here.  \citet{EWuyts14} find the [NII]/H$\alpha$
ratios are constant for all but clump \emph{e}, indicating a constant metallicity\footnote{As noted 
earlier, clump \emph{e} suffers from contribution to the line emission from outflows and
a higher measured electron density, and we therefore suspect that the higher [NII]/H$\alpha$ ratio does not indicate
a higher metallicity.  Moreover, \citet{EWuyts14} find that the $R_{23}$ metallicity 
indicator \citep{Zaritsky94, Kobulnicky04} is consistent amongst all of the clumps.}.
We further assume that the region is radiation-bounded.  
At these relatively low densities and electron temperatures, the emission from
He~I~$\lambda5876$ is almost entirely recombination; collisional excitation
contributes $<10\%$ \citep{Porter09}.

One complication in the plasma modeling is that, in the low density limit, an infinite
number of energy levels of the H and He atoms would need to be treated
to correctly calculate the recombination line fluxes \citep{Bauman05, Porter05, Porter07}.  
However, each atom must be approximated with a finite number of energy levels.  
The default settings of Cloudy have insufficient levels to correctly
predict He~I/H~I recombination line ratios.  Therefore, we follow
\citet{Porter07}, who calculated theoretical He~I/H~I recombination
line ratios for the Orion nebula, by using 40 resolved levels and 60
collapsed levels for both the H~I and He~I atoms.  
Such models push high-end desktop computers to their limits; each 
Cloudy model took 3 days to run (versus 10 minutes for models with the
default number of levels for He and H).  Although we cannot formally prove
that this finite numerical approximation yields the correct answer,
we do find that the line ratios converge as the number of levels increases:  
increasing the number of resolved levels from 6 (default) to 40
changes the predicted He~I~$\lambda5876$ line ratio by $6\%$; increasing from
40 to 50 resolved levels changes the line ratio by only $0.5\%$.

Figure~\ref{fig:HeI_masscutoff} shows the theoretical value of the 
He~I $\lambda5876$/H$\beta$ line ratio (left panel), as computed
from our Cloudy models.  For continuous star formation with an Orion He abundance pattern, 30\% solar metalicity, 
density of 180 cm$^{-3}$,
and a \citet{Salpeter} IMF with an upper mass cutoff anywhere in the range of 60 to 120 M$_{\odot}$, the 
saturated He~I~$\lambda5876$/H$\beta$ ratio is 0.123. 

The saturated value of the ratio is sensitive to the He/H abundance ratio and the population of massive stars, 
and is remarkably insensitive to anything else.  The relationship with He abundance is linear: for example, 
increasing the relative He abundance by 10\% increases the He~I/H$\beta$ line ratio by 10\%.  The saturated value 
is insensitive to gas-phase metallicity; increasing the metallicity from 30\% to 40\% increases the saturated 
value by only 1.5\%.  The saturated value is insensitive to density: increasing the density from 180 cm$^{-3}$ to 600 cm$^{-3}$ 
increases the saturated value by only 5\%.  
We choose the Orion He abundance pattern (Y$_{\mathrm{p}}$=0.29)\footnote{Calculated from Table 7.2 from the Hazy 
documentation for Cloudy, \url{http://data.nublado.org/cloudy\_releases/c13/}},
as this is equivalent to the Solar He abundance and most useful for future reference.  As the
saturated ratio scales linearly with He abundance, the value we calculated can be adopted to other He/H abundances.

To explore the evolution of line ratios after star formation ceases, we assume an instantaneous starburst and age
the spectrum to compute both the He~I $\lambda5876$/H$\beta$ and He~II $\lambda4686$/H$\beta$ line ratios with time
(right panel of Figure~\ref{fig:HeI_masscutoff}).  The He~I $\lambda5876$/H$\beta$ line ratio steeply drops from the saturated 
value after 5 Myr, when the hottest stars die.  Similarly, the He~II $\lambda4686$/H$\beta$ line ratio, sensitive 
to very hard ultraviolet photons, drops after only 3 Myr.  In reality, there will be a more gradual evolution of these
line ratios as star formation is not truncated instantaneously.  Figure~\ref{fig:HeI_masscutoff} illustrates the general
trends for this simplified case of an instantaneous starburst.  

\subsubsection{Interpretation}

Figure~\ref{fig:HbHeI} shows that the He~I $\lambda5876$/H$\beta$ ratios for four knots are consistent with the saturated value, 
within the error bars.  
This implies that their ionizing spectra are similar to that of a \citet{Salpeter} present-day mass
function extending above 60 M$_{\odot}$.  The high He II $\lambda4686$/H$\beta$ and
He I $\lambda5876$/H$\beta$ line ratios for knot \emph{g} suggest that massive stars have formed within the last 3 Myr.

Two knots, \emph{e} and \emph{f}, have measured
He I $\lambda5876$/H$\beta$ ratios below the saturated value.  
\citet{Karczewski13} measure a similar ratio of 0.079 in a local irregular 
starburst galaxy NGC 4449, although earlier measurements 
by \citet{Martin97} of 4 local irregular galaxies found a lower ratio of only 0.01--0.02.  Interestingly,
these galaxies are considered to be starbursting in the local Universe, but they would be considered ``normal'' at $z=1.7$
due to the strong evolution in the SFRs of galaxies \citep[e.g., ][]{Whitaker12b}.  
If the low values we measure in this paper are not due to measurement errors, 
there are three straight forward interpretations of the lower He~I $\lambda5876$/H$\beta$ ratios measured 
in clumps \emph{e} and \emph{f}. The first explanation would be that those clumps have a lower Helium 
to Hydrogen abundance ratio. Alternatively, these clumps could be missing their most massive stars.  
Such a situation could arise if either the IMF has an upper mass cutoff such that the most massive stars never formed, 
or if the IMF extends to high masses, but enough time has passed that the most massive stars have died.  
The intermediate He~I $\lambda5876$/H$\beta$ line ratio value together with the low He~II $\lambda4686$/H$\beta$ 
line ratio of $0.03\pm0.02$ in knot \emph{e} suggest that star formation shut down about 5-–6 Myr ago 
(see Figure~\ref{fig:HeI_masscutoff}, right panel).

The latter interpretation fits in with our understanding of the merger history of RCS0327.
It is possible that all of the star-forming clumps are formed
from gas clouds near the established stellar population in knot \emph{g} as a result of the gravitational interaction.  
In this simple picture, those star-forming clumps physically closest to knot \emph{g} formed earliest and the clumps increasingly
further from knot \emph{g} formed at later times.
As knot \emph{u} is physically closest to knot \emph{g}, the gas reservoir may continue to be 
replenished even though this clump was likely the first to form, yet still has a high observed He~I/H$\beta$ ratio. 
On the other hand, knots \emph{f} and \emph{e} formed next in our
simple picture and their low He~I $\lambda5876$/H$\beta$ line ratios (and He~II $\lambda4686$/H$\beta$ for knot \emph{e}) imply 
that they have already depleted their gas supply and 
are no longer actively forming new stars.  Finally, knots \emph{d, c} and \emph{b} are the furthest in distance from knot \emph{g}
and the last to form their stars.  Given the high observed He~I $\lambda5876$/H$\beta$ line ratios, the most massive stars in
these star-forming regions must still be contributing to the ionizing spectrum.
Although we cannot formally constrain the absolute ages of the different star-forming knots with the
He to H line ratio diagnostic, we are able to set limits on the time since the last burst of new star formation.  This
diagnostic therefore provides indirect evidence for the order in which star formation is stopping in the individual star-forming
clumps.

\section{Discussion}
\label{sec:discussion}

The data presented herein enables the unique opportunity to study in great detail the physical properties of 
individual star-forming regions in the ongoing merger RCS0327 at $z=1.7$, the epoch marking the cosmic peak of star
formation activity.
Through a spatially-resolved kinematic analysis of the galaxy, \citet{EWuyts14} find evidence that the gas-rich knot \emph{g}
has an established
stellar population that interacted with an older, gas-poor and similarly massive stellar population.
Consistent with this scenario, \citet{EWuyts14} observe both a tidal arm extending west of knot \emph{g}, 
as well as chain of star-forming clumps extending east that are thought to have
formed from a stream of gas pulled away from knot \emph{g} as a result of the interaction (see Figure~\ref{fig:image}).

Here, we present the results from HST/WFC3 grism spectroscopy and place them in the context of our 
current understanding of star formation history of RCS0327.
We measure the spatial variations of the extinction in RCS0327 through the observed H$\gamma$/H$\beta$ line ratios, 
finding an average extinction of $E(B-V)_{gas}=0.40\pm0.07$.  When combined with the SED modeling analysis
of \citet{EWuyts14}, we find the ratio of the average extinction of the gas relative to the stars is 
$\mathrm{E(B-V)_{gas}/E(B-V)_{star}}=1.4\pm0.3$.
This value is consistent with the results in the literature \citep{Wuyts13, Price13, Kashino13}, 
who measure similar levels of extra
extinction towards the star-forming regions in distant galaxies.  

There is a marginal trend at the 2$\sigma$ significance level within the grism observations alone for spatial variations
in the extinction, with the slightly larger values measured with increased distance from the 
established stellar population of knot \emph{g}.  However, when including higher spectral resolution ground-based
Keck/NIRSPEC \citep{Rigby11} and Magellan/FIRE \citep{EWuyts14} ground-based
measurements, this trend is no longer statistically significant.  In this galaxy,
the extinction does not appear to have significant variations amongst the different star-forming regions, and a
global measurement should therefore adequately describe the properties of RCS0327.

Combining the [OIII]$\lambda$5007/H$\beta$ line ratio measurements from the WFC3 grism spectroscopy
with [NII]/H$\alpha$ line ratios from \citet{EWuyts14}, we find that the majority of the star-forming regions
in RCS0327 fall along the local abundance sequence in the BPT diagram, albeit extended into the low-metallicity 
regime that is not well populated in the local Universe.  
Consistent with our understanding of the star formation history, we expect the measured metallicities (as implied 
by the [NII]/H$\alpha$ line ratios presented in Figure~\ref{fig:bpt}) for these star-forming knots to be the 
same as their suspected gas reservoir source, knot \emph{g}.  We observe a higher [NII]/H$\alpha$ line ratio 
in knot \emph{e} alone, falling along the extreme abundance sequence for $z=1.7$ predicted by \citet{Kewley13}.  We attribute
this discrepency to the combination of a higher electron density \citep{EWuyts14} and shock excitation.
As no discrepancy exists for the $R_{23}$ metallicity indicators \citep{EWuyts14}, there is no strong evidence to 
suggest that this clump should have a different metallicity.  
\citet{Rich13} demonstrate that at the mid-to-end stages of a merger, low velocity shocks
have the largest contribution ($\sim$40--60\%) to the observed H$\alpha$ line flux.
For four clumps in RCS0327, \citet{EWuyts14} find that the broad underlying component in the H$\alpha$ emission line profile
contributes about 40\% of the total H$\alpha$ flux, on average.  Placing the \citet{Rich13} result in the context
of the BPT diagram, it is perhaps not surprising that 
knot \emph{e} is offset, but rather that the rest of the star-forming regions appear to have ``normal'' emission-line ratios
with little contamination from shocks.

The star formation in RCS0327 is enhanced as a result of the ongoing interaction, with measured SFRs derived
from extinction-corrected H$\beta$ line fluxes for seven
individually-resolved star forming clumps falling 1--2 dex above the observed star formation sequence.
These observations confirm previous theoretical and empirical results that galaxy 
mergers have enhanced SFRs relative to the ``normal'' star formation sequence.  

Merger simulations find that an AGN does not turn on until the final stages of a 
merger \citep[e.g.,][]{diMatteo05, Springel05, PHopkins06},
after the black holes have coalesced.  These findings are generally supported by observations \citep[e.g., ][]{Teng10}.
The stellar population is predicted to be detected as an aging
starburst with an age of several 100 Myr by the time the AGN becomes visible in the optical.
As RCS0327 is likely in the mid-to-end stages of a merger with two separated mass concentrations, 
knot \emph{g} and an older stellar population (see reconstructed image in Figure~\ref{fig:image}), we do not expect 
to find an AGN. There is no strong evidence for an actively accreting supermassive black hole in RCS0327:
the emission line ratios for all star-forming regions are consistent with the H~II-origin parameter space of the BPT diagram,
no point source was seen in Chandra imaging for the three star-forming regions closest to the likely nucleus, and
we do not observe variability between two epochs of HST images taken over a three month timescale. 

In this emerging picture, we expect the formation age of these star-forming knots to correspond to their 
physical distance from knot \emph{g}, the suspected source of the 
gas reservoir.  Although measuring the absolute age of the stellar populations is beyond this data set, 
the established stellar population in knot \emph{g} and the nearby knot \emph{u}, as well 
as the furthest (probably most recently formed) knots \emph{b, c} and \emph{d} have the high 
He~I $\lambda5876$/H$\beta$ recombination line ratios, consistent with the saturated value.  
Such high helium-to-hydrogen ratios imply that these star-forming regions are
dominated by hot O-stars.  Knot \emph{u} is physically close 
to knot \emph{g} and may therefore still be accreting gas to fuel ongoing star-formation.
As suggested by the lower He~I/H$\beta$ line ratio measurements (and the low He~II/H$\beta$ line 
ratio in knot \emph{e}), the most massive stars have already
died in knots \emph{f} and \emph{e}.  
Together with theoretical models, the He~I/H$\beta$ and He~II/H$\beta$ line ratio measurements imply that knot \emph{e} had to 
stop actively forming stars about 5--6 Myr ago, presumably due to the depletion of the 
gas reservoirs and stellar feedback.  Thus, we may be probing the decline of star formation in a distant 
galaxy on $\sim100$ pc scales, enabled by gravitational lensing and the high spatial resolution of HST.

There is a wealth of information to be learned through such exceptional lensed galaxies as RCS0327.  Through future
HST/WFC3 imaging and spectroscopic studies of distant lensed galaxies, we can map the spatial variations of the
extinction and star formation for larger samples to thereby better understand how distant galaxies formed their stars.

\begin{acknowledgements}
The authors wish to acknowledge Alaina Henry for insightful discussions,
and Alejo Stark for his help with the AGN analysis.
We thank the anonymous referee for useful comments and a                                                                                   
careful reading of the paper.
KEW is supported by an appointment to the NASA
Postdoctoral Program at the Goddard Space Flight Center,
administered by Oak Ridge Associated Universities through a 
contract with NASA.
This research is based on observations made with the NASA/ESA Hubble Space 
Telescope, obtained at the Space Telescope Science Institute, which is operated by the 
Association of Universities for Research in Astronomy, Inc., under NASA contract NAS 5-26555. 
These observations are associated with program 12726.  Support for program number 12726 was 
provided by NASA through a grant from the Space Telescope Science Institute, which is operated 
by the Association of Universities for Research in Astronomy, Inc., under NASA contract NAS5-26555.
We thank Gary Ferland and Claus Leitherer for making the 
Cloudy and Starbust99 tools publicly available.
\end{acknowledgements}

\addcontentsline{toc}{chapter}{\numberline {}{\sc References}}


\begin{thebibliography}{111}
\expandafter\ifx\csname natexlab\endcsname\relax\def\natexlab#1{#1}\fi

\bibitem[{{Abazajian} {et~al.}(2009){Abazajian}, {Adelman-McCarthy},
  {Ag{\"u}eros}, {Allam}, {Allende Prieto}, {An}, {Anderson}, {Anderson},
  {Annis}, {Bahcall}, \& et~al.}]{Abazajian09}
{Abazajian}, K.~N., {Adelman-McCarthy}, J.~K., {Ag{\"u}eros}, M.~A., {et~al.}
  2009, \apjs, 182, 543

\bibitem[{{Aird} {et~al.}(2010){Aird}, {Nandra}, {Laird}, {Georgakakis},
  {Ashby}, {Barmby}, {Coil}, {Huang}, {Koekemoer}, {Steidel}, \&
  {Willmer}}]{Aird10}
{Aird}, J., {Nandra}, K., {Laird}, E.~S., {et~al.} 2010, \mnras, 401, 2531

\bibitem[{{Baldwin} {et~al.}(1981){Baldwin}, {Phillips}, \&
  {Terlevich}}]{Baldwin81}
{Baldwin}, J.~A., {Phillips}, M.~M., \& {Terlevich}, R. 1981, \pasp, 93, 5

\bibitem[{{Bauer} {et~al.}(2013){Bauer}, {Hopkins}, {Gunawardhana}, {Taylor},
  {Baldry}, {Bamford}, {Bland-Hawthorn}, {Brough}, {Brown}, {Cluver},
  {Colless}, {Conselice}, {Croom}, {Driver}, {Foster}, {Jones}, {Lara-Lopez},
  {Liske}, {L{\'o}pez-S{\'a}nchez}, {Loveday}, {Norberg}, {Owers}, {Pimbblet},
  {Robotham}, {Sansom}, \& {Sharp}}]{Bauer13}
{Bauer}, A.~E., {Hopkins}, A.~M., {Gunawardhana}, M., {et~al.} 2013, \mnras,
  434, 209

\bibitem[{{Bashkin} \& {Stoner}(1975)}]{Bashkin75}
{Bashkin}, S., \& {Stoner}, J.~O. 1975, {Atomic energy levels and Grotrian
  Diagrams - Vol.1: Hydrogen I - Phosphorus XV; Vol.2: Sulfur I - Titanium
  XXII}

\bibitem[{{Bauman} {et~al.}(2005){Bauman}, {Porter}, {Ferland}, \&
  {MacAdam}}]{Bauman05}
{Bauman}, R.~P., {Porter}, R.~L., {Ferland}, G.~J., {et~al.} 2005, \apj, 628,
  541

\bibitem[{{Bell} \& {de Jong}(2000)}]{Bell00}
{Bell}, E.~F., \& {de Jong}, R.~S. 2000, \mnras, 312, 497

\bibitem[{{Bell} {et~al.}(2012){Bell}, {van der Wel}, {Papovich}, {Kocevski},
  {Lotz}, {McIntosh}, {Kartaltepe}, {Faber}, {Ferguson}, {Koekemoer}, {Grogin},
  {Wuyts}, {Cheung}, {Conselice}, {Dekel}, {Dunlop}, {Giavalisco},
  {Herrington}, {Koo}, {McGrath}, {de Mello}, {Rix}, {Robaina}, \&
  {Williams}}]{Bell12}
{Bell}, E.~F., {van der Wel}, A., {Papovich}, C., {et~al.} 2012, \apj, 753, 167

\bibitem[{{Bournaud} {et~al.}(2011){Bournaud}, {Chapon}, {Teyssier}, {Powell},
  {Elmegreen}, {Elmegreen}, {Duc}, {Contini}, {Epinat}, \&
  {Shapiro}}]{Bournaud11}
{Bournaud}, F., {Chapon}, D., {Teyssier}, R., {et~al.} 2011, \apj, 730, 4

\bibitem[{{Brammer} {et~al.}(2012{\natexlab{a}}){Brammer},
  {S{\'a}nchez-Janssen}, {Labb{\'e}}, {da Cunha}, {Erb}, {Franx}, {Fumagalli},
  {Lundgren}, {Marchesini}, {Momcheva}, {Nelson}, {Patel}, {Quadri}, {Rix},
  {Skelton}, {Schmidt}, {van der Wel}, {van Dokkum}, {Wake}, \&
  {Whitaker}}]{Brammer12b}
{Brammer}, G.~B., {S{\'a}nchez-Janssen}, R., {Labb{\'e}}, I., {et~al.}
  2012{\natexlab{a}}, \apjl, 758, L17

\bibitem[{{Brammer} {et~al.}(2012{\natexlab{b}}){Brammer}, {van Dokkum},
  {Franx}, {Fumagalli}, {Patel}, {Rix}, {Skelton}, {Kriek}, {Nelson},
  {Schmidt}, {Bezanson}, {da Cunha}, {Erb}, {Fan}, {F{\"o}rster Schreiber},
  {Illingworth}, {Labb{\'e}}, {Leja}, {Lundgren}, {Magee}, {Marchesini},
  {McCarthy}, {Momcheva}, {Muzzin}, {Quadri}, {Steidel}, {Tal}, {Wake},
  {Whitaker}, \& {Williams}}]{Brammer12}
{Brammer}, G.~B., {van Dokkum}, P.~G., {Franx}, M., {et~al.}
  2012{\natexlab{b}}, \apjs, 200, 13

\bibitem[{{Brinchmann} {et~al.}(2004){Brinchmann}, {Charlot}, {White},
  {Tremonti}, {Kauffmann}, {Heckman}, \& {Brinkmann}}]{Brinchmann04}
{Brinchmann}, J., {Charlot}, S., {White}, S.~D.~M., {et~al.} 2004, \mnras, 351,
  1151

\bibitem[{{Calzetti} {et~al.}(2000){Calzetti}, {Armus}, {Bohlin}, {Kinney},
  {Koornneef}, \& {Storchi-Bergmann}}]{Calzetti00}
{Calzetti}, D., {Armus}, L., {Bohlin}, R.~C., {et~al.} 2000, \apj, 533, 682

\bibitem[{{Cassata} {et~al.}(2013){Cassata}, {Giavalisco}, {Williams}, {Guo},
  {Lee}, {Renzini}, {Ferguson}, {Faber}, {Barro}, {McIntosh}, {Lu}, {Bell},
  {Koo}, {Papovich}, {Ryan}, {Conselice}, {Grogin}, {Koekemoer}, \&
  {Hathi}}]{Cassata13}
{Cassata}, P., {Giavalisco}, M., {Williams}, C.~C., {et~al.} 2013, \apj, 775,
  106

\bibitem[{{Chabrier}(2003)}]{Chabrier}
{Chabrier}, G. 2003, \pasp, 115, 763

\bibitem[{{Cullen} {et~al.}(2013){Cullen}, {Cirasuolo}, {McLure}, \&
  {Dunlop}}]{Cullen13}
{Cullen}, F., {Cirasuolo}, M., {McLure}, R.~J., {et~al.} 2013, ArXiv e-prints

\bibitem[{{Daddi} {et~al.}(2007){Daddi}, {Dickinson}, {Morrison}, {Chary},
  {Cimatti}, {Elbaz}, {Frayer}, {Renzini}, {Pope}, {Alexander}, {Bauer},
  {Giavalisco}, {Huynh}, {Kurk}, \& {Mignoli}}]{Daddi07}
{Daddi}, E., {Dickinson}, M., {Morrison}, G., {et~al.} 2007, \apj, 670, 156

\bibitem[{{Dav{\'e}} {et~al.}(2012){Dav{\'e}}, {Finlator}, \&
  {Oppenheimer}}]{Dave12}
{Dav{\'e}}, R., {Finlator}, K., \& {Oppenheimer}, B.~D. 2012, \mnras, 421, 98

\bibitem[{{Di Matteo} {et~al.}(2005){Di Matteo}, {Springel}, \&
  {Hernquist}}]{diMatteo05}
{Di Matteo}, T., {Springel}, V., \& {Hernquist}, L. 2005, \nat, 433, 604

\bibitem[{{Dom{\'{\i}}nguez} {et~al.}(2013){Dom{\'{\i}}nguez}, {Siana},
  {Henry}, {Scarlata}, {Bedregal}, {Malkan}, {Atek}, {Ross}, {Colbert},
  {Teplitz}, {Rafelski}, {McCarthy}, {Bunker}, {Hathi}, {Dressler}, {Martin},
  \& {Masters}}]{Dominguez13}
{Dom{\'{\i}}nguez}, A., {Siana}, B., {Henry}, A.~L., {et~al.} 2013, \apj, 763,
  145

\bibitem[{{Dopita} \& {Evans}(1986)}]{Dopita86}
{Dopita}, M.~A., \& {Evans}, I.~N. 1986, \apj, 307, 431

\bibitem[{{Dutton} {et~al.}(2010){Dutton}, {van den Bosch}, \&
  {Dekel}}]{Dutton10}
{Dutton}, A.~A., {van den Bosch}, F.~C., \& {Dekel}, A. 2010, \mnras, 405, 1690

\bibitem[{{Elbaz} {et~al.}(2007){Elbaz}, {Daddi}, {Le Borgne}, {Dickinson},
  {Alexander}, {Chary}, {Starck}, {Brandt}, {Kitzbichler}, {MacDonald},
  {Nonino}, {Popesso}, {Stern}, \& {Vanzella}}]{Elbaz07}
{Elbaz}, D., {Daddi}, E., {Le Borgne}, D., {et~al.} 2007, \aap, 468, 33

\bibitem[{{Erb} {et~al.}(2006){Erb}, {Steidel}, {Shapley}, {Pettini}, {Reddy},
  \& {Adelberger}}]{Erb06}
{Erb}, D.~K., {Steidel}, C.~C., {Shapley}, A.~E., {et~al.} 2006, \apj, 647, 128

\bibitem[{{Ferland} {et~al.}(2013){Ferland}, {Porter}, {van Hoof}, {Williams},
  {Abel}, {Lykins}, {Shaw}, {Henney}, \& {Stancil}}]{Ferland13}
{Ferland}, G.~J., {Porter}, R.~L., {van Hoof}, P.~A.~M., {et~al.} 2013, RMxAA,
  49, 137

\bibitem[{{Ferreras} {et~al.}(2005){Ferreras}, {Lisker}, {Carollo}, {Lilly}, \&
  {Mobasher}}]{Ferreras05}
{Ferreras}, I., {Lisker}, T., {Carollo}, C.~M., {et~al.} 2005, \apj, 635, 243

\bibitem[{{Foreman-Mackey} {et~al.}(2013){Foreman-Mackey}, {Hogg}, {Lang}, \&
  {Goodman}}]{ForemanMackey13}
{Foreman-Mackey}, D., {Hogg}, D.~W., {Lang}, D., {et~al.} 2013, \pasp, 125, 306

\bibitem[{{F{\"o}rster Schreiber} {et~al.}(2009){F{\"o}rster Schreiber},
  {Genzel}, {Bouch{\'e}}, {Cresci}, {Davies}, {Buschkamp}, {Shapiro},
  {Tacconi}, {Hicks}, {Genel}, {Shapley}, {Erb}, {Steidel}, {Lutz},
  {Eisenhauer}, {Gillessen}, {Sternberg}, {Renzini}, {Cimatti}, {Daddi},
  {Kurk}, {Lilly}, {Kong}, {Lehnert}, {Nesvadba}, {Verma}, {McCracken},
  {Arimoto}, {Mignoli}, \& {Onodera}}]{ForsterSchreiber09}
{F{\"o}rster Schreiber}, N.~M., {Genzel}, R., {Bouch{\'e}}, N., {et~al.} 2009,
  \apj, 706, 1364

\bibitem[{{F{\"o}rster Schreiber} {et~al.}(2001){F{\"o}rster Schreiber},
  {Genzel}, {Lutz}, {Kunze}, \& {Sternberg}}]{ForsterSchreiber01}
{F{\"o}rster Schreiber}, N.~M., {Genzel}, R., {Lutz}, D., {et~al.} 2001, \apj,
  552, 544

\bibitem[{{Garn} {et~al.}(2010){Garn}, {Sobral}, {Best}, {Geach}, {Smail},
  {Cirasuolo}, {Dalton}, {Dunlop}, {McLure}, \& {Farrah}}]{Garn10}
{Garn}, T., {Sobral}, D., {Best}, P.~N., {et~al.} 2010, \mnras, 402, 2017

\bibitem[{{Gies} {et~al.}(2002){Gies}, {Penny}, {Mayer}, {Drechsel}, \&
  {Lorenz}}]{Gies02}
{Gies}, D.~R., {Penny}, L.~R., {Mayer}, P., {et~al.} 2002, \apj, 574, 957

\bibitem[{{Gonz{\'a}lez} {et~al.}(2010){Gonz{\'a}lez}, {Labb{\'e}}, {Bouwens},
  {Illingworth}, {Franx}, {Kriek}, \& {Brammer}}]{Gonzalez10}
{Gonz{\'a}lez}, V., {Labb{\'e}}, I., {Bouwens}, R.~J., {et~al.} 2010, \apj,
  713, 115

\bibitem[{{Groves} {et~al.}(2012){Groves}, {Brinchmann}, \&
  {Walcher}}]{Groves12}
{Groves}, B., {Brinchmann}, J., \& {Walcher}, C.~J. 2012, \mnras, 419, 1402

\bibitem[{{Guo} {et~al.}(2013){Guo}, {Zheng}, \& {Fu}}]{KGuo13}
{Guo}, K., {Zheng}, X.~Z., \& {Fu}, H. 2013, \apj, 778, 23

\bibitem[{{Hainline} {et~al.}(2009){Hainline}, {Shapley}, {Kornei}, {Pettini},
  {Buckley-Geer}, {Allam}, \& {Tucker}}]{Hainline09}
{Hainline}, K.~N., {Shapley}, A.~E., {Kornei}, K.~A., {et~al.} 2009, \apj, 701,
  52

\bibitem[{{Hopkins} \& {Beacom}(2006)}]{Hopkins06}
{Hopkins}, A.~M., \& {Beacom}, J.~F. 2006, \apj, 651, 142

\bibitem[{{Hopkins} {et~al.}(2006){Hopkins}, {Hernquist}, {Cox}, {Di Matteo},
  {Robertson}, \& {Springel}}]{PHopkins06}
{Hopkins}, P.~F., {Hernquist}, L., {Cox}, T.~J., {et~al.} 2006, \apjs, 163, 1

\bibitem[{{Hopkins} {et~al.}(2008){Hopkins}, {Hernquist}, {Cox}, \& {Kere{\v
  s}}}]{PHopkins08}
---. 2008, \apjs, 175, 356

\bibitem[{{Jogee} {et~al.}(2009){Jogee}, {Miller}, {Penner}, {Skelton},
  {Conselice}, {Somerville}, {Bell}, {Zheng}, {Rix}, {Robaina}, {Barazza},
  {Barden}, {Borch}, {Beckwith}, {Caldwell}, {Peng}, {Heymans}, {McIntosh},
  {H{\"a}u{\ss}ler}, {Jahnke}, {Meisenheimer}, {Sanchez}, {Wisotzki}, {Wolf},
  \& {Papovich}}]{Jogee09}
{Jogee}, S., {Miller}, S.~H., {Penner}, K., {et~al.} 2009, \apj, 697, 1971

\bibitem[{{Jones} {et~al.}(2010){Jones}, {Swinbank}, {Ellis}, {Richard}, \&
  {Stark}}]{Jones10}
{Jones}, T.~A., {Swinbank}, A.~M., {Ellis}, R.~S., {et~al.} 2010, \mnras, 404,
  1247

\bibitem[{{Karczewski} {et~al.}(2013){Karczewski}, {Barlow}, {Page}, {Kuin},
  {Ferreras}, {Baes}, {Bendo}, {Boselli}, {Cooray}, {Cormier}, {De Looze},
  {Galametz}, {Galliano}, {Lebouteiller}, {Madden}, {Pohlen}, {R{\'e}my-Ruyer},
  {Smith}, \& {Spinoglio}}]{Karczewski13}
{Karczewski}, O.~{\L}., {Barlow}, M.~J., {Page}, M.~J., {et~al.} 2013, \mnras,
  431, 2493

\bibitem[{{Kashino} {et~al.}(2013){Kashino}, {Silverman}, {Rodighiero},
  {Renzini}, {Arimoto}, {Daddi}, {Lilly}, {Sanders}, {Kartaltepe}, {Zahid},
  {Nagao}, {Sugiyama}, {Capak}, {Carollo}, {Chu}, {Hasinger}, {Ilbert},
  {Kajisawa}, {Kewley}, {Koekemoer}, {Kova{\v c}}, {Le F{\`e}vre}, {Masters},
  {McCracken}, {Onodera}, {Scoville}, {Strazzullo}, {Symeonidis}, \&
  {Taniguchi}}]{Kashino13}
{Kashino}, D., {Silverman}, J.~D., {Rodighiero}, G., {et~al.} 2013, \apjl, 777,
  L8

\bibitem[{{Kauffmann} {et~al.}(2003){Kauffmann}, {Heckman}, {Tremonti},
  {Brinchmann}, {Charlot}, {White}, {Ridgway}, {Brinkmann}, {Fukugita}, {Hall},
  {Ivezi{\'c}}, {Richards}, \& {Schneider}}]{Kauffmann03b}
{Kauffmann}, G., {Heckman}, T.~M., {Tremonti}, C., {et~al.} 2003, \mnras, 346,
  1055

\bibitem[{{Kennicutt}(1998)}]{Kennicutt98}
{Kennicutt}, Jr., R.~C. 1998, \araa, 36, 189

\bibitem[{{Kennicutt} {et~al.}(2000){Kennicutt}, {Bresolin}, {French}, \&
  {Martin}}]{Kennicutt00}
{Kennicutt}, Jr., R.~C., {Bresolin}, F., {French}, H., {et~al.} 2000, \apj,
  537, 589

\bibitem[{{Kewley} {et~al.}(2013){Kewley}, {Dopita}, {Leitherer}, {Dave},
  {Yuan}, {Allen}, {Groves}, \& {Sutherland}}]{Kewley13}
{Kewley}, L.~J., {Dopita}, M.~A., {Leitherer}, C., {et~al.} 2013, ArXiv
  e-prints

\bibitem[{{Kewley} {et~al.}(2001){Kewley}, {Dopita}, {Sutherland}, {Heisler},
  \& {Trevena}}]{Kewley01}
{Kewley}, L.~J., {Dopita}, M.~A., {Sutherland}, R.~S., {et~al.} 2001, \apj,
  556, 121

\bibitem[{{Kobulnicky} \& {Kewley}(2004)}]{Kobulnicky04}
{Kobulnicky}, H.~A., \& {Kewley}, L.~J. 2004, \apj, 617, 240

\bibitem[{{Kriek} \& {Conroy}(2013)}]{Kriek13}
{Kriek}, M., \& {Conroy}, C. 2013, \apjl, 775, L16

\bibitem[{{Kroupa}(2001)}]{Kroupa01}
{Kroupa}, P. 2001, \mnras, 322, 231

\bibitem[{{Leitherer} {et~al.}(2010){Leitherer}, {Ortiz Ot{\'a}lvaro},
  {Bresolin}, {Kudritzki}, {Lo Faro}, {Pauldrach}, {Pettini}, \&
  {Rix}}]{Leitherer10}
{Leitherer}, C., {Ortiz Ot{\'a}lvaro}, P.~A., {Bresolin}, F., {et~al.} 2010,
  \apjs, 189, 309

\bibitem[{{Leitherer} {et~al.}(1999){Leitherer}, {Schaerer}, {Goldader},
  {Gonz{\'a}lez Delgado}, {Robert}, {Kune}, {de Mello}, {Devost}, \&
  {Heckman}}]{Starburst99}
{Leitherer}, C., {Schaerer}, D., {Goldader}, J.~D., {et~al.} 1999, \apjs, 123,
  3

\bibitem[{{Leitner}(2012)}]{Leitner12}
{Leitner}, S.~N. 2012, \apj, 745, 149

\bibitem[{{Lotz} {et~al.}(2008){Lotz}, {Jonsson}, {Cox}, \& {Primack}}]{Lotz08}
{Lotz}, J.~M., {Jonsson}, P., {Cox}, T.~J., {et~al.} 2008, \mnras, 391, 1137

\bibitem[{{MacArthur} {et~al.}(2004){MacArthur}, {Courteau}, {Bell}, \&
  {Holtzman}}]{MacArthur04}
{MacArthur}, L.~A., {Courteau}, S., {Bell}, E., {et~al.} 2004, \apjs, 152, 175

\bibitem[{{Magdis} {et~al.}(2012){Magdis}, {Daddi}, {B{\'e}thermin}, {Sargent},
  {Elbaz}, {Pannella}, {Dickinson}, {Dannerbauer}, {da Cunha}, {Walter},
  {Rigopoulou}, {Charmandaris}, {Hwang}, \& {Kartaltepe}}]{Magdis12}
{Magdis}, G.~E., {Daddi}, E., {B{\'e}thermin}, M., {et~al.} 2012, \apj, 760, 6

\bibitem[{{Magdis} {et~al.}(2010){Magdis}, {Rigopoulou}, {Huang}, \&
  {Fazio}}]{Magdis10}
{Magdis}, G.~E., {Rigopoulou}, D., {Huang}, J.-S., {et~al.} 2010, \mnras, 401,
  1521

\bibitem[{{Mancini} {et~al.}(2011){Mancini}, {F{\"o}rster Schreiber},
  {Renzini}, {Cresci}, {Hicks}, {Peng}, {Vergani}, {Lilly}, {Carollo},
  {Pozzetti}, {Zamorani}, {Daddi}, {Genzel}, {Maraston}, {McCracken},
  {Tacconi}, {Bouch{\'e}}, {Davies}, {Oesch}, {Shapiro}, {Mainieri}, {Lutz},
  {Mignoli}, \& {Sternberg}}]{Mancini11}
{Mancini}, C., {F{\"o}rster Schreiber}, N.~M., {Renzini}, A., {et~al.} 2011,
  \apj, 743, 86

\bibitem[{{Marchesini} {et~al.}(2009){Marchesini}, {van Dokkum}, {F{\"o}rster
  Schreiber}, {Franx}, {Labb{\'e}}, \& {Wuyts}}]{Marchesini09}
{Marchesini}, D., {van Dokkum}, P.~G., {F{\"o}rster Schreiber}, N.~M., {et~al.}
  2009, \apj, 701, 1765

\bibitem[{{Martin} \& {Kennicutt}(1997)}]{Martin97}
{Martin}, C.~L., \& {Kennicutt}, Jr., R.~C. 1997, \apj, 483, 698

\bibitem[{{Mihos} \& {Hernquist}(1994)}]{Mihos94}
{Mihos}, J.~C., \& {Hernquist}, L. 1994, \apjl, 431, L9

\bibitem[{{Newman} {et~al.}(2014){Newman}, {Buschkamp}, {Genzel}, {F{\"o}rster
  Schreiber}, {Kurk}, {Sternberg}, {Gnat}, {Rosario}, {Mancini}, {Lilly},
  {Renzini}, {Burkert}, {Carollo}, {Cresci}, {Davies}, {Eisenhauer}, {Genel},
  {Shapiro Griffin}, {Hicks}, {Lutz}, {Naab}, {Peng}, {Tacconi}, {Wuyts},
  {Zamorani}, {Vergani}, \& {Weiner}}]{SNewman14}
{Newman}, S.~F., {Buschkamp}, P., {Genzel}, R., {et~al.} 2014, \apj, 781, 21

\bibitem[{{Niemela} \& {Bassino}(1994)}]{Niemela94}
{Niemela}, V.~S., \& {Bassino}, L.~P. 1994, \apj, 437, 332

\bibitem[{{Noeske} {et~al.}(2007){Noeske}, {Weiner}, {Faber}, {Papovich},
  {Koo}, {Somerville}, {Bundy}, {Conselice}, {Newman}, {Schiminovich}, {Le
  Floc'h}, {Coil}, {Rieke}, {Lotz}, {Primack}, {Barmby}, {Cooper}, {Davis},
  {Ellis}, {Fazio}, {Guhathakurta}, {Huang}, {Kassin}, {Martin}, {Phillips},
  {Rich}, {Small}, {Willmer}, \& {Wilson}}]{Noeske07a}
{Noeske}, K.~G., {Weiner}, B.~J., {Faber}, S.~M., {et~al.} 2007, \apjl, 660,
  L43

\bibitem[{{Oey} {et~al.}(2000){Oey}, {Dopita}, {Shields}, \& {Smith}}]{Oey00}
{Oey}, M.~S., {Dopita}, M.~A., {Shields}, J.~C., {et~al.} 2000, \apjs, 128, 511

\bibitem[{{Osterbrock} \& {Ferland}(2006)}]{OsterbrockFerland}
{Osterbrock}, D.~E., \& {Ferland}, G.~J. 2006, {Astrophysics of gaseous nebulae
  and active galactic nuclei}

\bibitem[{{Pannella} {et~al.}(2009){Pannella}, {Carilli}, {Daddi}, {McCracken},
  {Owen}, {Renzini}, {Strazzullo}, {Civano}, {Koekemoer}, {Schinnerer},
  {Scoville}, {Smol{\v c}i{\'c}}, {Taniguchi}, {Aussel}, {Kneib}, {Ilbert},
  {Mellier}, {Salvato}, {Thompson}, \& {Willott}}]{Pannella09}
{Pannella}, M., {Carilli}, C.~L., {Daddi}, E., {et~al.} 2009, \apjl, 698, L116

\bibitem[{{Patel} {et~al.}(2011){Patel}, {Kelson}, {Holden}, {Franx}, \&
  {Illingworth}}]{Patel11}
{Patel}, S.~G., {Kelson}, D.~D., {Holden}, B.~P., {et~al.} 2011, \apj, 735, 53

\bibitem[{{Peng} {et~al.}(2010){Peng}, {Lilly}, {Kova{\v c}}, {Bolzonella},
  {Pozzetti}, {Renzini}, {Zamorani}, {Ilbert}, {Knobel}, {Iovino}, {Maier},
  {Cucciati}, {Tasca}, {Carollo}, {Silverman}, {Kampczyk}, {de Ravel},
  {Sanders}, {Scoville}, {Contini}, {Mainieri}, {Scodeggio}, {Kneib}, {Le
  F{\`e}vre}, {Bardelli}, {Bongiorno}, {Caputi}, {Coppa}, {de la Torre},
  {Franzetti}, {Garilli}, {Lamareille}, {Le Borgne}, {Le Brun}, {Mignoli},
  {Perez Montero}, {Pello}, {Ricciardelli}, {Tanaka}, {Tresse}, {Vergani},
  {Welikala}, {Zucca}, {Oesch}, {Abbas}, {Barnes}, {Bordoloi}, {Bottini},
  {Cappi}, {Cassata}, {Cimatti}, {Fumana}, {Hasinger}, {Koekemoer},
  {Leauthaud}, {Maccagni}, {Marinoni}, {McCracken}, {Memeo}, {Meneux}, {Nair},
  {Porciani}, {Presotto}, \& {Scaramella}}]{Peng10}
{Peng}, Y.-j., {Lilly}, S.~J., {Kova{\v c}}, K., {et~al.} 2010, \apj, 721, 193

\bibitem[{{Porter} {et~al.}(2005){Porter}, {Bauman}, {Ferland}, \&
  {MacAdam}}]{Porter05}
{Porter}, R.~L., {Bauman}, R.~P., {Ferland}, G.~J., {et~al.} 2005, \apjl, 622,
  L73

\bibitem[{{Porter} {et~al.}(2007){Porter}, {Ferland}, \& {MacAdam}}]{Porter07}
{Porter}, R.~L., {Ferland}, G.~J., \& {MacAdam}, K.~B. 2007, \apj, 657, 327

\bibitem[{{Porter} {et~al.}(2009){Porter}, {Ferland}, {MacAdam}, \&
  {Storey}}]{Porter09}
{Porter}, R.~L., {Ferland}, G.~J., {MacAdam}, K.~B., {et~al.} 2009, \mnras,
  393, L36

\bibitem[{{Price} {et~al.}(2013){Price}, {Kriek}, {Brammer}, {Conroy}, {Forster
  Schreiber}, {Franx}, {Fumagalli}, {Lundgren}, {Momcheva}, {Nelson}, {Rix},
  {Skelton}, {van Dokkum}, {Whitaker}, \& {Wuyts}}]{Price13}
{Price}, S.~H., {Kriek}, M., {Brammer}, G.~B., {et~al.} 2013, ArXiv e-prints

\bibitem[{{Reddy} {et~al.}(2010){Reddy}, {Erb}, {Pettini}, {Steidel}, \&
  {Shapley}}]{Reddy10}
{Reddy}, N.~A., {Erb}, D.~K., {Pettini}, M., {et~al.} 2010, \apj, 712, 1070

\bibitem[{{Rich} {et~al.}(2010){Rich}, {Dopita}, {Kewley}, \& {Rupke}}]{Rich10}
{Rich}, J.~A., {Dopita}, M.~A., {Kewley}, L.~J., {et~al.} 2010, \apj, 721, 505

\bibitem[{{Rich} {et~al.}(2013){Rich}, {Kewley}, \& {Dopita}}]{Rich13}
{Rich}, J.~A., {Kewley}, L.~J., \& {Dopita}, M.~A. 2013, ArXiv e-prints

\bibitem[{{Rich} {et~al.}(2012){Rich}, {Torrey}, {Kewley}, {Dopita}, \&
  {Rupke}}]{Rich12}
{Rich}, J.~A., {Torrey}, P., {Kewley}, L.~J., {et~al.} 2012, \apj, 753, 5

\bibitem[{{Richard} {et~al.}(2011){Richard}, {Jones}, {Ellis}, {Stark},
  {Livermore}, \& {Swinbank}}]{Richard11}
{Richard}, J., {Jones}, T., {Ellis}, R., {et~al.} 2011, \mnras, 413, 643

\bibitem[{{Rigby} \& {Rieke}(2004)}]{Rigby04}
{Rigby}, J.~R., \& {Rieke}, G.~H. 2004, \apj, 606, 237

\bibitem[{{Rigby} {et~al.}(2011){Rigby}, {Wuyts}, {Gladders}, {Sharon}, \&
  {Becker}}]{Rigby11}
{Rigby}, J.~R., {Wuyts}, E., {Gladders}, M.~D., {et~al.} 2011, \apj, 732, 59

\bibitem[{{Rodighiero} {et~al.}(2011){Rodighiero}, {Daddi}, {Baronchelli},
  {Cimatti}, {Renzini}, {Aussel}, {Popesso}, {Lutz}, {Andreani}, {Berta},
  {Cava}, {Elbaz}, {Feltre}, {Fontana}, {F{\"o}rster Schreiber},
  {Franceschini}, {Genzel}, {Grazian}, {Gruppioni}, {Ilbert}, {Le Floch},
  {Magdis}, {Magliocchetti}, {Magnelli}, {Maiolino}, {McCracken}, {Nordon},
  {Poglitsch}, {Santini}, {Pozzi}, {Riguccini}, {Tacconi}, {Wuyts}, \&
  {Zamorani}}]{Rodighiero11}
{Rodighiero}, G., {Daddi}, E., {Baronchelli}, I., {et~al.} 2011, \apjl, 739,
  L40

\bibitem[{{Salpeter}(1955)}]{Salpeter}
{Salpeter}, E.~E. 1955, \apj, 121, 161

\bibitem[{{Sanchez-Blazquez} {et~al.}(2013){Sanchez-Blazquez},
  {Rosales-Ortega}, {Diaz}, \& {Sanchez}}]{SanchezBlazquez13}
{Sanchez-Blazquez}, P., {Rosales-Ortega}, F., {Diaz}, A., {et~al.} 2013, ArXiv
  e-prints

\bibitem[{{Sharon} {et~al.}(2012){Sharon}, {Gladders}, {Rigby}, {Wuyts},
  {Koester}, {Bayliss}, \& {Barrientos}}]{Sharon12}
{Sharon}, K., {Gladders}, M.~D., {Rigby}, J.~R., {et~al.} 2012, \apj, 746, 161

\bibitem[{{Speagle} {et~al.}(2014){Speagle}, {Steinhardt}, {Capak}, \&
  {Silverman}}]{Speagle14}
{Speagle}, J.~S., {Steinhardt}, C.~L., {Capak}, P.~L., {et~al.} 2014, ArXiv
  e-prints

\bibitem[{{Springel} {et~al.}(2005){Springel}, {Di Matteo}, \&
  {Hernquist}}]{Springel05}
{Springel}, V., {Di Matteo}, T., \& {Hernquist}, L. 2005, \mnras, 361, 776

\bibitem[{{Stark} {et~al.}(2013){Stark}, {Auger}, {Belokurov}, {Jones},
  {Robertson}, {Ellis}, {Sand}, {Moiseev}, {Eagle}, \& {Myers}}]{Stark13}
{Stark}, D.~P., {Auger}, M., {Belokurov}, V., {et~al.} 2013, \mnras, 436, 1040

\bibitem[{{Stecher}(1965)}]{Stecher65}
{Stecher}, T.~P. 1965, \apj, 142, 1683

\bibitem[{{Szomoru} {et~al.}(2013){Szomoru}, {Franx}, {van Dokkum}, {Trenti},
  {Illingworth}, {Labb{\'e}}, \& {Oesch}}]{Szomoru13}
{Szomoru}, D., {Franx}, M., {van Dokkum}, P.~G., {et~al.} 2013, \apj, 763, 73

\bibitem[{{Takata} {et~al.}(2006){Takata}, {Sekiguchi}, {Smail}, {Chapman},
  {Geach}, {Swinbank}, {Blain}, \& {Ivison}}]{Takata06}
{Takata}, T., {Sekiguchi}, K., {Smail}, I., {et~al.} 2006, \apj, 651, 713

\bibitem[{{Teng} \& {Veilleux}(2010)}]{Teng10}
{Teng}, S.~H., \& {Veilleux}, S. 2010, \apj, 725, 1848

\bibitem[{{Teplitz} {et~al.}(2000){Teplitz}, {McLean}, {Becklin}, {Figer},
  {Gilbert}, {Graham}, {Larkin}, {Levenson}, \& {Wilcox}}]{Teplitz00}
{Teplitz}, H.~I., {McLean}, I.~S., {Becklin}, E.~E., {et~al.} 2000, \apjl, 533,
  L65

\bibitem[{{Tomczak} {et~al.}(2013){Tomczak}, {Quadri}, {Tran}, {Labbe},
  {Straatman}, {Papovich}, {Glazebrook}, {Allen}, {Kacprzak},
  {Kawinwanichakij}, {Kelson}, {McCarthy}, {Mehrtens}, {Monson}, {Persson},
  {Spitler}, {Tilvi}, \& {van Dokkum}}]{Tomczak13}
{Tomczak}, A.~R., {Quadri}, R.~F., {Tran}, K.-V.~H., {et~al.} 2013, ArXiv
  e-prints

\bibitem[{{van Dokkum}(2005)}]{vanDokkum05}
{van Dokkum}, P.~G. 2005, \aj, 130, 2647

\bibitem[{{Vanzi} {et~al.}(1996){Vanzi}, {Rieke}, {Martin}, \&
  {Shields}}]{Vanzi96}
{Vanzi}, L., {Rieke}, G.~H., {Martin}, C.~L., {et~al.} 1996, \apj, 466, 150

\bibitem[{{V{\'a}zquez} \& {Leitherer}(2005)}]{Vazquez05}
{V{\'a}zquez}, G.~A., \& {Leitherer}, C. 2005, \apj, 621, 695

\bibitem[{{Veilleux} \& {Osterbrock}(1987)}]{Veilleux87}
{Veilleux}, S., \& {Osterbrock}, D.~E. 1987, \apjs, 63, 295

\bibitem[{{Weingartner} \& {Draine}(2001)}]{Weingartner01}
{Weingartner}, J.~C., \& {Draine}, B.~T. 2001, \apj, 548, 296

\bibitem[{{Weisz} {et~al.}(2011){Weisz}, {Dalcanton}, {Williams}, {Gilbert},
  {Skillman}, {Seth}, {Dolphin}, {McQuinn}, {Gogarten}, {Holtzman}, {Rosema},
  {Cole}, {Karachentsev}, \& {Zaritsky}}]{Weisz11}
{Weisz}, D.~R., {Dalcanton}, J.~J., {Williams}, B.~F., {et~al.} 2011, \apj,
  739, 5

\bibitem[{{Whitaker} {et~al.}(2011){Whitaker}, {Labb{\'e}}, {van Dokkum},
  {Brammer}, {Kriek}, {Marchesini}, {Quadri}, {Franx}, {Muzzin}, {Williams},
  {Bezanson}, {Illingworth}, {Lee}, {Lundgren}, {Nelson}, {Rudnick}, {Tal}, \&
  {Wake}}]{Whitaker11}
{Whitaker}, K.~E., {Labb{\'e}}, I., {van Dokkum}, P.~G., {et~al.} 2011, \apj,
  735, 86

\bibitem[{{Whitaker} {et~al.}(2012){Whitaker}, {van Dokkum}, {Brammer}, \&
  {Franx}}]{Whitaker12b}
{Whitaker}, K.~E., {van Dokkum}, P.~G., {Brammer}, G., {et~al.} 2012, \apjl,
  754, L29

\bibitem[{{Wuyts} {et~al.}(2010){Wuyts}, {Barrientos}, {Gladders}, {Sharon},
  {Bayliss}, {Carrasco}, {Gilbank}, {Yee}, {Koester}, \&
  {Mu{\~n}oz}}]{EWuyts10}
{Wuyts}, E., {Barrientos}, L.~F., {Gladders}, M.~D., {et~al.} 2009, \apj, 724,
  1182

\bibitem[{{Wuyts} {et~al.}(2014){Wuyts}, {Rigby}, {Gladders}, \&
  {Sharon}}]{EWuyts14}
{Wuyts}, E., {Rigby}, J.~R., {Gladders}, M.~D., {et~al.} 2014, \apj, 781, 61

\bibitem[{{Wuyts} {et~al.}(2013){Wuyts}, {F{\"o}rster Schreiber}, {Nelson},
  {van Dokkum}, {Brammer}, {Chang}, {Faber}, {Ferguson}, {Franx}, {Fumagalli},
  {Genzel}, {Grogin}, {Kocevski}, {Koekemoer}, {Lundgren}, {Lutz}, {McGrath},
  {Momcheva}, {Rosario}, {Skelton}, {Tacconi}, {van der Wel}, \&
  {Whitaker}}]{Wuyts13}
{Wuyts}, S., {F{\"o}rster Schreiber}, N.~M., {Nelson}, E.~J., {et~al.} 2013,
  \apj, 779, 135

\bibitem[{{Wuyts} {et~al.}(2011){Wuyts}, {F{\"o}rster Schreiber}, {van der
  Wel}, {Magnelli}, {Guo}, {Genzel}, {Lutz}, {Aussel}, {Barro}, {Berta},
  {Cava}, {Graci{\'a}-Carpio}, {Hathi}, {Huang}, {Kocevski}, {Koekemoer},
  {Lee}, {Le Floc'h}, {McGrath}, {Nordon}, {Popesso}, {Pozzi}, {Riguccini},
  {Rodighiero}, {Saintonge}, \& {Tacconi}}]{Wuyts11b}
{Wuyts}, S., {F{\"o}rster Schreiber}, N.~M., {van der Wel}, A., {et~al.} 2011,
  \apj, 742, 96

\bibitem[{{Yabe} {et~al.}(2014){Yabe}, {Ohta}, {Iwamuro}, {Akiyama}, {Tamura},
  {Yuma}, {Kimura}, {Takato}, {Moritani}, {Sumiyoshi}, {Maihara}, {Silverman},
  {Dalton}, {Lewis}, {Bonfield}, {Lee}, {Curtis-Lake}, {Macaulay}, \&
  {Clarke}}]{Yabe14}
{Yabe}, K., {Ohta}, K., {Iwamuro}, F., {et~al.} 2014, \mnras, 437, 3647

\bibitem[{{Yabe} {et~al.}(2012){Yabe}, {Ohta}, {Iwamuro}, {Yuma}, {Akiyama},
  {Tamura}, {Kimura}, {Takato}, {Moritani}, {Sumiyoshi}, {Maihara},
  {Silverman}, {Dalton}, {Lewis}, {Bonfield}, {Lee}, {Curtis Lake}, {Macaulay},
  \& {Clarke}}]{Yabe12}
---. 2012, \pasj, 64, 60

\bibitem[{{Yeh} \& {Matzner}(2012)}]{Yeh12}
{Yeh}, S.~C.~C., \& {Matzner}, C.~D. 2012, \apj, 757, 108

\bibitem[{{Yoshikawa} {et~al.}(2010){Yoshikawa}, {Akiyama}, {Kajisawa},
  {Alexander}, {Ohta}, {Suzuki}, {Tokoku}, {Uchimoto}, {Konishi}, {Yamada},
  {Tanaka}, {Omata}, {Nishimura}, {Koekemoer}, {Brandt}, \&
  {Ichikawa}}]{Yoshikawa10}
{Yoshikawa}, T., {Akiyama}, M., {Kajisawa}, M., {et~al.} 2010, \apj, 718, 112

\bibitem[{{Yuan} {et~al.}(2012){Yuan}, {Kewley}, {Swinbank}, \&
  {Richard}}]{Yuan12}
{Yuan}, T.-T., {Kewley}, L.~J., {Swinbank}, A.~M., {et~al.} 2012, \apj, 759, 66

\bibitem[{{Yuan} {et~al.}(2011){Yuan}, {Kewley}, {Swinbank}, {Richard}, \&
  {Livermore}}]{Yuan11}
---. 2011, \apjl, 732, L14

\bibitem[{{Zahid} {et~al.}(2013{\natexlab{a}}){Zahid}, {Kashino}, {Silverman},
  {Kewley}, {Daddi}, {Renzini}, {Rodighiero}, {Nagao}, {Arimoto}, {Sanders},
  {Kartaltepe}, {Lilly}, {Maier}, {Capak}, {Carollo}, {Chu}, {Hasinger},
  {Ilbert}, {Kajisawa}, {Koekemoer}, {Kovac}, {Le Fevre}, {Masters},
  {McCracken}, {Onodera}, {Scoville}, {Strazzullo}, {Sugiyama}, {Taniguchi}, \&
  {The COSMOS Team}}]{Zahid13b}
{Zahid}, H.~J., {Kashino}, D., {Silverman}, J.~D., {et~al.} 2013{\natexlab{a}},
  ArXiv e-prints

\bibitem[{{Zahid} {et~al.}(2013{\natexlab{b}}){Zahid}, {Yates}, {Kewley}, \&
  {Kudritzki}}]{Zahid13}
{Zahid}, H.~J., {Yates}, R.~M., {Kewley}, L.~J., {et~al.} 2013{\natexlab{b}},
  \apj, 763, 92

\bibitem[{{Zaritsky} {et~al.}(1994){Zaritsky}, {Kennicutt}, \&
  {Huchra}}]{Zaritsky94}
{Zaritsky}, D., {Kennicutt}, Jr., R.~C., \& {Huchra}, J.~P. 1994, \apj, 420, 87

\end{thebibliography}

\end{document}